\documentclass[11pt]{article}
\usepackage{graphicx}
\usepackage{amsmath}
\usepackage{mathtools}
\usepackage{appendix}
\usepackage{tikz}
\usepackage{subcaption}
\usepackage{textcomp, gensymb}

\usepackage[affil-it]{authblk}
\usepackage[numbers,sort&compress]{natbib}
\usepackage[margin=1in]{geometry}
\usepackage{setspace}
\usepackage{pdflscape}
\usepackage[indent=20pt]{parskip}
\usepackage{makecell}

\begin{document}

\title{Illustrating implications of misaligned causal questions and statistics in settings with competing events and interest in treatment mechanisms}
\author{Takuya Kawahara$^{1,2}$, 
Sean McGrath$^{1,3}$, and 
Jessica G. Young$^{1,4}$\\
$^{1}$Department of Population Medicine, Harvard Medical School and \\Harvard Pilgrim Health Care Institute, Boston, MA, U.S.A. \\
$^{2}$Clinical Research Promotion Center, The University of Tokyo Hospital, Tokyo, Japan \\
$^{3}$Department of Biostatistics, Yale School of Public Health, New Haven, CT, U.S.A. \\
$^{4}$Department of Epidemiology, Harvard T. H. Chan School of Public Health, Boston, MA, U.S.A.}
\date{\today}

\maketitle
\begin{abstract}
   In the presence of competing events, many investigators are interested in a direct treatment effect on the event of interest that does not capture treatment effects on competing events. Classical survival analysis methods that treat competing events like censoring events, at best, target a controlled direct effect: the effect of the treatment under a difficult to imagine and typically clinically irrelevant scenario where competing events are somehow eliminated.   A separable direct effect, quantifying the effect of a future modified version of the treatment, is an alternative direct effect notion that may better align with an investigator's underlying causal question. In this paper, we provide insights into the implications of naively applying an estimator constructed for a controlled direct effect (i.e., ``censoring by competing events'') when the actual causal effect of interest is a separable direct effect. We illustrate the degree to which controlled and separable direct effects may take different values, possibly even different signs, and the degree to which these two different effects may be differentially impacted by violation and/or near violation of their respective identifying conditions under a range of data generating scenarios. Finally, we provide an empirical comparison of inverse probability of censoring weighting to an alternative weighted estimator specifically structured for a separable effect using data from a randomized trial of estrogen therapy and prostate cancer mortality.
\end{abstract}

\maketitle

\section{Introduction}
Randomized trials are considered the ``gold standard'' for causal inference because, in principle, they recover a \textsl{total} treatment effect on a specified study outcome via all of its causal mechanisms \citep{robins1992identifiability}. However, in settings with competing events, specific mechanisms making up a total treatment effect may be of consequential interest to investigators \citep{young2020causal}.

For example, consider a historical trial of estrogen therapy (versus placebo) in men diagnosed with prostate cancer, with study outcome ``death due to prostate cancer'' \citep{prostatedata}.  Some men in both arms died of cardiovascular-related events.  Cardiovascular death is a competing event for prostate cancer death because, once an individual dies of a cardiovascular-related event, he is prevented from subsequently dying of prostate cancer.  Stating the outcome of interest as prostate cancer-specific death (as opposed to all-cause mortality) inherently suggests interest in one hypothesized mechanism by which estrogen affects survival (via slowing/stopping prostate cancer progression).  However, the total effect on prostate cancer death may capture other mechanisms via competing events (such that individuals are ``protected'' by first dying of cardiovascular events) \citep{dobbs2019estrogens}. 

For investigators who  desire an effect capturing isolated treatment mechanism(s), an emerging literature on separable effects \citep{robins2010alternative, didelez2019defining, stensrud2022marginal, stensrud2021generalized, robins2022interventionist} -- effects of new versions of the study treatment under hypothesized modifications -- may be relevant. In this case, modified treatment effects might be defined capturing only mechanisms that ``directly'' affect the event of interest, not via the treatment's effect on competing events; i.e. ``separable direct effects'' \citep{stensrud2022marginal}.  

The default approach to analyzing competing events data is to ``censor'' an individual's follow-up upon experiencing a competing event and then proceed with statistics for right-censored data \citep{geskus2016data}. Certain implementations of this approach might be interpreted in terms of an effect under ``elimination of competing events'', coinciding with a case of a ``controlled direct effect'' formalized in the causal mediation literature \citep{robins1992identifiability,young2020causal}.  However, such effects refer to scenarios that are typically inconceivable.  By referring to impossible scenarios, either now or in the future, such controlled direct effects will typically not have interpretable clinical implications nor align with what clinical investigators want to know when this is transparently articulated.  

 Despite recent statements to the contrary \citep{austin2025inverse}, a controlled direct effect and separable direct effect are different effects that can take different values.  Thus, even in a circumstance where an investigator has estimated a controlled direct effect ``very well'', they may have estimated the effect of actual interest ``very poorly'' if that is a separable direct effect.  One source of confusion may be that nonparametric identifying conditions for both effects share superficial similarities such as the requirement that there are no unmeasured shared causes of the competing event and event of interest at any time \citep{young2020causal,stensrud2022marginal}. Yet, there are underappreciated differences in the respective required identifying conditions for these different effects and differences in the impact of failure of these conditions on statistics designed to estimate them, respectively.   

Premised on a scenario where the actual study question aligns with a separable direct effect, here we distinguish three compounding sources of ``error'' in an analysis of competing events data: (1) ``estimand error'', an interpretational error arising from formalizing the causal estimand in a way that is misaligned with the actual study question; (2) ``non-identification error'', structural (``causal'') bias due to failure of nonparametric identifying conditions; and (3)``statistical error'', bias and variance of an estimator relative to only an observed data target (without reference to a causal question or model). We compare these error sources under a range of data generating scenarios -- including varying the frequency of competing events, the existence of an unmeasured common cause of competing events and the event of interest, and the presence of ``near'' positivity violations - across two types of weighted estimators: the widely used inverse probability of censoring weighted (IPCW) estimator where competing events are treated as censoring events and a weighted estimator structured with a separable direct effect in mind \citep{stensrud2022marginal}.  We end with a practical comparison of these two estimators in a re-analysis of data from the estrogen therapy  trial.  

\section{Background\label{causalstory}}

In this section we will briefly review and compare formal definitions of a controlled direct and a separable direct effect for competing events settings and conditions for their identification, with full technical details found elsewhere \citep{young2020causal,stensrud2022marginal,stensrud2021generalized}. Suppose Figure 1a is a causal directed acyclic graph (DAG) \citep{robins2010alternative} depicting underlying data generating assumptions on a randomized trial.

\begin{figure}[t]
\begin{tabular}{cc}
\begin{minipage}[b]{.45\textwidth}
    \centering
    \begin{tikzpicture}
      \node (A) at (0,0) {$A$};
      \node (D1) at (2,0) {$D$};
      \node (Y1) at (4,0) {$Y$};
      \node (L) at (3,1.5) {$L$};
      \node (U) at (3,-1.2) {$U$};
      \draw[->] (A) edge (D1);
      \draw[->] (A) edge [bend right=30] (Y1);
      \draw[->] (D1) edge (Y1);
      \draw[->] (L) edge (D1);
      \draw[->] (L) edge (Y1);
      \draw[->] (U) edge [dashed] (D1);
      \draw[->] (U) edge [dashed] (Y1);
    \end{tikzpicture}
  \subcaption{}
\end{minipage}
&
\begin{minipage}[b]{.45\textwidth}
\centering
    \begin{tikzpicture}
      \node (A) at (0,0) {$A|a$};
      \node (D1) at (2,0) {$D^a$};
      \node (Y1) at (4,0) {$Y^a$};
      \node (L) at (3,1.5) {$L$};
      \node (U) at (3,-1.2) {$U$};
      \draw[->] (A) edge (D1);
      \draw[->] (A) edge [bend right=30] (Y1);
      \draw[->] (D1) edge (Y1);
      \draw[->] (L) edge (D1);
      \draw[->] (L) edge (Y1);
      \draw[->] (U) edge [dashed] (D1);
      \draw[->] (U) edge [dashed] (Y1);
    \end{tikzpicture}
   \subcaption{}
\end{minipage}
\end{tabular}

\caption{Causal diagrams implicitly (a) versus explicitly (b) communicating the underlying counterfactual causal model in which the total effect is defined and interpreted. The arrow from $D$ to $Y$ (or $D^a$ to $Y^a$) is by definition present when $D$ is a competing event for $Y$.  The additional arrow from $A$ to $D$ in (a) and correspondingly $a$ to $D^a$ in (b) communicates that investigators do not rule out that estrogen therapy may prevent prostate cancer death via causing death in other ways.}
\label{fig:total effect}
\end{figure}
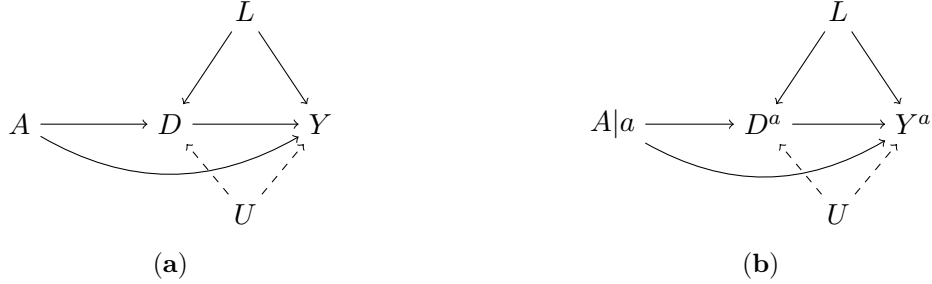

Let $A$ denote randomized baseline treatment assignment (e.g. $A=1$ assignment to estrogen therapy, $A=0$ placebo), $Y$ an indicator of failure from an event of interest (e.g. death due to prostate cancer) and $D$ an indicator of any competing event (e.g. cardiovascular-related death) by an end of follow-up time of interest (e.g. 50 months). For simplicity, but without meaningful loss of generality of core arguments, we suppress the time-varying nature of the event of interest and competing event processes, assume $D$ temporally precedes $Y$, and assume perfect adherence to baseline randomization. Throughout, we will also assume that there are no common causes of $D$ and $Y$ affected by $A$.  None of these simplifications are required for defining or identifying any effects considered below \citep{young2020causal,stensrud2021generalized}.  The arrow from $D$ to $Y$ in Figure 1a depicts the dependence inherent to competing events: an individual is prevented from dying of prostate cancer once they have died from a cardiovascular-related event. The causal model in Figure 1a allows that there may exist pre-treatment measured common causes ($L$) of the event of interest and competing event. $U$ denotes an unmeasured pre-treatment variable.  In subsequent sections, we will vary assumptions on the presence or absence of the dashed arrows in Figure 1a. 

The paths consisting of only right directed arrows (causal paths) connecting $A$ to $Y$ in Figure 1a implicitly comprise the total effect of $A$ on $Y$: they depict \textit{all} assumed possible mechanisms by which $A$ affects $Y$ \citep{robins1992identifiability}.  This is more explicit in Figure 1b, a single world intervention graph (SWIG) \citep{richardson2013single}, depicting  counterfactual outcomes defining a total effect: a contrast in $\Pr[Y^{a}=1]$ for $a=1$ versus $a=0$, relative to this model \citep{young2020causal}, where $Y^a$ denotes an individual's (possibly counterfactual) event of interest status under an intervention where $A$ is set to $a$.  By Figure 1, the total effect is identified in this trial by a contrast in $\Pr[Y=1|A=a]$,
a simple function of the measured variables \citep{young2020causal, hernan2020causal}. 

However, in general, the total effect does not align with the effect of interest for an investigator who wishes to isolate only the treatment's mechanisms that operate on the event of interest \textsl{not} via the determinism created by competing events \citep{young2020causal}.  The misalignment of the total effect with such a goal under the working causal model in Figure 1 is represented implicitly by the path $A\rightarrow D\rightarrow Y$ in Figure 1a and more explicitly by $a\rightarrow D^a\rightarrow Y^a$ in Figure 1b. As reviewed above, the arrow from the competing event to the event of interest will always be there when competing events exist, regardless of the nature of treatment.  However, if subject matter background justified removal of the arrows in Figure 1 from treatment to competing events, then the total effect would in fact align with such a goal (Figure 2a).  

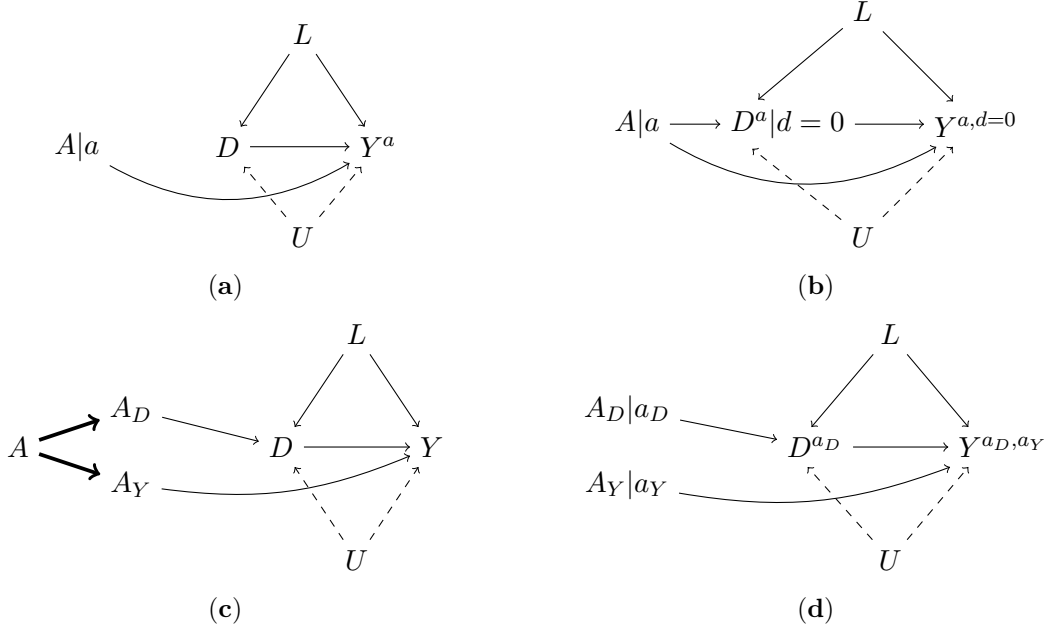
\begin{figure}[t]
\begin{tabular}{cc}
\begin{minipage}[b]{.45\textwidth}
\centering
    \begin{tikzpicture}
      \node (A) at (0,0) {$A|a$};
      \node (D1) at (2,0) {$D$};
      \node (Y1) at (4,0) {$Y^a$};
      \node (L) at (3,1.5) {$L$};
      \node (U) at (3,-1.2) {$U$};
      \draw[->] (A) edge [bend right=30] (Y1);
      \draw[->] (D1) edge (Y1);
      \draw[->] (L) edge (D1);
      \draw[->] (L) edge (Y1);
      \draw[->] (U) edge [dashed] (D1);
      \draw[->] (U) edge [dashed] (Y1);
    \end{tikzpicture}
    \subcaption{}
\end{minipage}
&
\begin{minipage}[b]{.45\textwidth}
\centering
  \begin{tikzpicture}
      \node (A) at (0,0) {$A|a$};
      \node (D1) at (2,0) {$D^a|d=0$};
      \node (Y1) at (4.5,0) {$Y^{a, d=0}$};
      \node (L) at (3,1.5) {$L$};
      \node (U) at (3,-1.5) {$U$};
      \draw[->] (A) edge (D1);
      \draw[->] (A) edge [bend right=30] (Y1);
      \draw[->] (L) edge (D1.140);
      \draw[->] (L) edge (Y1);
      \draw[->] (D1) edge (Y1);
      \draw[->] (U) edge [dashed] (D1.215);
      \draw[->] (U) edge [dashed] (Y1);
    \end{tikzpicture}
    \subcaption{}
\end{minipage}\\

\begin{minipage}[b]{.45\textwidth}
  \centering
    \begin{tikzpicture}
      \node (A) at (-2, 0) {$A$};
      \node (AD) at (-0.5,0.5) {$A_D$};
      \node (AY) at (-0.5,-0.5) {$A_Y$};
      \node (D1) at (1.5,0) {$D$};
      \node (Y1) at (3.5,0) {$Y$};
      \node (L) at (2.5,1.5) {$L$};
      \node (U) at (2.5,-1.5) {$U$};
      \draw[->, line width=0.5mm] (A) edge (AD);
      \draw[->, line width=0.5mm] (A) edge (AY);
      \draw[->] (AD) edge (D1);
      \draw[->] (AY) edge [bend right=15] (Y1);
      \draw[->] (D1) edge (Y1);
      \draw[->] (L) edge (D1);
      \draw[->] (L) edge (Y1);
      \draw[->] (U) edge [dashed] (D1);
      \draw[->] (U) edge [dashed] (Y1);
    \end{tikzpicture}
\subcaption{}
\end{minipage}
&
\begin{minipage}[b]{.45\textwidth}
    \centering
    \begin{tikzpicture}
      \node (AD) at (-1,0.5) {$A_D|a_D$};
      \node (AY) at (-1,-0.5) {$A_Y|a_Y$};
      \node (D1) at (1.5,0) {$D^{a_D}$};
      \node (Y1) at (4,0) {$Y^{a_D, a_Y}$};
      \node (L) at (2.5,1.5) {$L$};
      \node (U) at (2.5,-1.5) {$U$};
      \draw[->] (AD) edge (D1);
      \draw[->] (AY) edge [bend right=15] (Y1);
      \draw[->] (D1) edge (Y1);
      \draw[->] (L) edge (D1.100);
      \draw[->] (L) edge (Y1.150);
      \draw[->] (U) edge [dashed] (D1.250);
      \draw[->] (U) edge [dashed] (Y1.210);
    \end{tikzpicture}
    \subcaption{}
\end{minipage}
\end{tabular}

\caption{Different notions of direct effect defined by different counterfactual causal models implicitly underlying Figure 1a. (a) A single world intervention graph \citep{richardson2013single} transformation of a restricted case of Figure 1a (with the arrow from $A$ to $D$ removed) under interventions that force the study treatment to value $a$. (b) A single world intervention graph transformation of Figure 1a under a joint intervention forcing $A$ to $a$ and $D^a$ to $0$. (c) An extended version of Figure 1a depicting the assumption that $(A_Y, A_D)$ is a decomposition of $A$ under full isolation \citep{robins2010alternative}. (d) A single world intervention graph transformation of the extended diagram in (c) under an intervention that jointly forces $A_Y$ to value $a_Y$ and $A_D$ to value $a_D$ \citep{stensrud2021generalized}.}
\label{fig:direct effects}
\end{figure}

\subsection{A controlled direct effect}

Investigators who do not wish to limit inference to the total effect must formalize another notion of a causal effect that isolates the mechanisms of interest under their causal model.  Let $Y^{a, d=0}$ denote an individual's counterfactual outcome under an additional intervention that (somehow) prevents all competing events.  In turn, the following is a case of a controlled direct effect \citep{robins1992identifiability} 
\begin{align}
  \mbox{CDE}_{0} \equiv \Pr[Y^{a=1, d=0}=1]-\Pr[Y^{a=0, d=0}=1] \label{cdecontrast},
\end{align}
which does not capture a treatment effect on the event of interest ``via competing events'' even when removing an arrow from $A$ to $D$ cannot be justified.  Provided that this effect notion is actually what motivates the study investigator, statistical methods for estimating a cumulative incidence/risk for ``censored'' data might then be justified and competing events can be correctly classified as ``censoring events'' \citep{young2020causal}.  The SWIG in Figure 2b is a transformation of the causal DAG in Figure 1a that more explicitly defines $\Pr(Y^{a, d=0}=1)$ and allows reasoning about its identification. By this model, we can write $\Pr(Y^{a, d=0}=1)$ in terms of the following function of factual \textsl{measured and unmeasured} variables 
\begin{align}
   \psi_0(a,d=0)=  \sum_{l,u} \Pr(Y=1|A=a, D=0, L=l, U=u) \Pr(L=l, U=u). \label{cdefactual}
\end{align}
provided this function is defined. In the restricted case where either of the dashed arrows in Figure 1a are absent, the controlled direct effect further reduces to the following function of \textsl{only measured} variables in our scenario
\begin{align}
 \mbox{CDE}_{obs} \equiv \tilde\psi(a=1,d=0)-\tilde\psi(a=0,d=0)  \label{cdeobscontrast}  
\end{align}
such that, for 
\begin{equation}
  \tilde{\pi}(A, L)\equiv\Pr(D=1|A,L),\label{psdeath}  
\end{equation}
we have
\begin{align}
    \tilde\psi(a,d=0) &= \sum_{l} \Pr(Y=1|A=a, D=0, L=l) \Pr(L=l) \nonumber\\
    &=E\left.\left[Y\frac{I(D=0)}{\{1-\tilde{\pi}(A=a, L)\}}\right|A=a\right],\label{cdeobs}
\end{align} 
provided it is defined.  In a study where $A$ is randomized, this is ensured by the following positivity condition with respect to competing events \citep{young2020causal} 
\begin{equation}
\Pr(A=a,L=l)>0\implies \{1- \tilde{\pi}(a, l)\}>0\mbox{ }a\in\{0,1\}.\label{pos}
\end{equation}
This positivity condition requires that there is no joint level of treatment and covariates such that everyone experiences the competing event.   

\subsection{A separable direct effect}
Despite the wide availability of statistics for the weighted outcome mean \eqref{cdeobs}, the controlled direct effect references a generally impossible scenario where competing events are universally eliminated.  In turn, this effect notion will often be misaligned with the investigators' intent when the goal is to inform clinical decision making.  \citet{stensrud2022marginal} posed an alternative effect definition that may in some cases directly align with an investigator's implicit target effect.  This will be the case when the underlying story is actually about the effect of a future modification to the study treatment. Specifically, let $A_Y$ and $A_D$ be two candidate treatments.  Suppose that, based on subject matter knowledge, the investigator believes the following \textsl{modified treatment assumption} is reasonable: receiving $A=1$ is equivalent (in terms of future counterfactual outcomes) to receiving $A_Y=A_D=1$, and receiving $A=0$ is equivalent to receiving $A_Y=A_D=0$ \citep{stensrud2021generalized}. One scenario where we expect this assumption to hold is when the candidate treatments are a physical decomposition of $A$ \citep{robins2010alternative}; however, it can be considered for candidates that are physically unrelated to the original study treatment \citep{stensrud2022marginal,stensrud2021generalized}. Separable effects generally can be defined as effects of joint interventions on these candidate treatments.  Under the modified treatment assumption relating these to the study treatment $A$, separable effects may isolate particular treatment mechanisms of interest to the investigator. Also see Lok and Bosch's work on the related notion of \textsl{organic effects} \citep{organiclok}.  

Specifically, Figure 2c is an extended version of Figure 1a, explicitly illustrating the special case of a decomposition assumption on $A$ \citep{robins2010alternative}, with $A_Y$ and $A_D$ (sets of) components. The bolded arrows depict the determinism in this case that if an individual receives $A=1$ in the trial they necessarily receive components $A_D=A_Y=1$ and receipt of $A=0$ implies both $A_D=A_Y=0$ \citep{robins2010alternative}. The SWIG in Figure 2d is a more explicit representation of the counterfactual causal model in which effects of joint interventions on these candidate treatments are defined. Figures 2c and 2d depict the assumption of \textsl{full isolation} \citep{stensrud2021generalized}, which jointly states: (1) there are no causal paths connecting treatment $A_Y$ and the event of interest $Y$ intersected by competing events and (2) there are no causal paths connecting treatment $A_D$ and competing events $D$ intersected by the event of interest $Y$. In this sense, under the modified treatment assumption and full isolation, an effect of $A_Y$ on $Y$ holding $A_D$ fixed captures a  (separable) direct effect of $A$ on $Y$ \citep{stensrud2021generalized, stensrud2022marginal}. Here we limit consideration to interest in separable effects under full isolation; however, weaker (partial) isolation conditions can also be considered \citep{stensrud2021generalized}. 

Let $Y^{a_Y,a_D}$ denote an individual's counterfactual outcome under an intervention jointly setting $A_Y=a_Y,A_D=a_D$, $a_Y\in{0,1},a_D\in{0,1}$ such that the following defines a separable direct effect of $A$ on $Y$ under Figure 2d and the modified treatment assumption \citep{stensrud2021generalized, stensrud2022marginal}
\begin{align}
  \mbox{SDE}^{a_D}_{0} \equiv \Pr(Y^{a_Y=1, a_D}=1)-\Pr(Y^{a_Y=0, a_D}=1)
  \label{sdecontrast}
\end{align}
for $a_D\in\{0,1\}$. There are, therefore, two separable direct effects, one indexed by $a_D=1$ and one by $a_D=0$.  The choice of $a_D$ of primary  interest to an investigator will depend on what motivates the investigator and the nature of $A_D$ \citep{stensrud2021generalized,stensrud2022marginal,stensrud2023conditional}.

Given the causal model depicted by Figure 2d is correct under a modified treatment assumption via a decomposition as in Figure 2c or otherwise, we can write $\Pr(Y^{a_Y, a_D}=1)$ in terms of the following function of factual \textsl{measured and unmeasured} variables in our scenario 
\begin{align}
\psi_0(a_Y, a_D)= &\sum_{l,u}\Pr(Y=1|A=a_Y, D=0, L=l, U=u) \times \nonumber\\
&\Pr(D=0|A=a_D, L=l, U=u) \Pr(L=l, U=u) \label{sdefactual}
\end{align} 
provided this function is defined \citep{stensrud2021generalized,stensrud2022marginal}. In the restricted case where either of the dashed arrows in Figure 1a are absent, the separable direct effect further reduces to the following function of factual \textsl{measured} variables in our scenario
\begin{align}
 \mbox{SDE}^{a_D}_{obs} \equiv \tilde\psi(a_Y=1,a_D)-\tilde\psi(a_Y=0,a_D)  \label{sdeobscontrast} 
\end{align}
such that, using definition \eqref{psdeath}
\begin{align}
    \tilde\psi(a_Y,a_D) &= \sum_{l}\Pr(Y=1|A=a_Y, D=0, L=l) 
    \Pr(D=0|A=a_D, L=l) \Pr(L=l) \nonumber\\
    &=E\left.\left[Y\frac{\{1-\tilde{\pi}(a_D, L)\}}{\{1-\tilde{\pi}(a_Y, L)\}}\right|A=a_Y\right]\label{sdeobs}
\end{align}
provided it is defined.  For a selected choice of $a_D$, this is ensured in a study where $A$ is physically randomized by a positivity condition with respect to competing events that can be written as follows: 
\begin{align}
&\Pr(A=a,L=l)>0\mbox{ and } \{1- \tilde{\pi}(a_D, l)\}>0\nonumber\\
&\implies \{1- \tilde{\pi}(a, l)\}>0\mbox{ }a\in\{0,1\}.\label{possde}
\end{align}
The condition \eqref{possde} is weaker than the positivity condition for the controlled direct effect \eqref{pos} in that it allows that there may exist a joint level of treatment and covariates such that everyone experiences the competing event if that joint level applies to treatment level $a_D$.

\section{Distinguishing error sources in two inverse probability weighted estimators }\label{errordefs}
We now illustrate different sources of error that can induce incorrect conclusions from an analysis of the observed data $(L,A,D,Y)$ when the analyst ``censors by competing events'' but interest is actually in a separable direct effect. 

\subsection{Inverse probability of censoring weighted estimator }
The following is an inverse probability of \textsl{censoring} weighted (IPCW) estimator with individuals experiencing the competing event (those with $D=1$) classified as \textsl{censored}:
\begin{align}
\widehat{\mbox{CDE}}_{obs} = \hat E\left.\left[Y\frac{I(D=0)}{\{1-\tilde{\pi}(A=1, L;\hat{\tilde{\beta}})\}}\right|A=1\right]-\hat E\left.\left[Y\frac{I(D=0)}{\{1-\tilde{\pi}(A=0, L;\hat{\tilde{\beta}})\}}\right|A=0\right] \label{cdehat}
\end{align}
with $\hat E$ the sample average operator, $\tilde{\pi}(A, L;\tilde{\beta})$ a parametric model for \eqref{psdeath} indexed by parameter vector $\tilde{\beta}$, and $\hat{\tilde{\beta}}$ the MLE of $\tilde{\beta}$ \citep{young2020causal}. 

We will refer to the effect that is truly of underlying interest to the investigator as the \textsl{actual causal target} of the analysis. Given our premise that the actual causal target is a \textsl{separable direct effect} \eqref{sdecontrast}, the total bias of an analysis that relies on this IPCW estimator \eqref{cdehat} is $\mbox{Bias}(\widehat{\mbox{CDE}}_{obs})_{SDE^{a_D}_0}=E[\widehat{\mbox{CDE}}_{obs}]- \mbox{SDE}^{a_D}_0$ and can be decomposed as follows: 
\begin{align}
\mbox{Bias}(\widehat{\mbox{CDE}}_{obs})_{SDE^{a_D}_0}&=\{\mbox{CDE}
_0-\mbox{SDE}^{a_D}_0\}\mbox{ } (\mbox{estimand error})\nonumber\\
&\quad+\{\mbox{CDE}_{obs}-\mbox{CDE}_0\}\mbox{ } (\mbox{non-identification error})\nonumber\\
&\quad+\{E[\widehat{\mbox{CDE}}_{obs}]-\mbox{CDE}_{obs}\} (\mbox{statistical error}).
  \label{biascdeobs}
\end{align}

The first component of bias captures \textit{estimand error}, quantifying the difference between the \textsl{actual causal target}, here a separable direct effect, and the \textsl{ostensible causal target}, the expectation of the estimator in the absence of the remaining two bias components. For the estimator \eqref{cdehat}, the ostensible causal target is the controlled direct effect \eqref{cdecontrast}. The second component of bias in \eqref{biascdeobs} 
captures \textit{non-identification error}, quantifying the difference between the \textsl{ostensible causal target} and the \textsl{statistical target}, the expectation of the estimator in the absence of the last bias component. We refer to the last bias component in \eqref{biascdeobs}, along with the variance  
\begin{equation}\mbox{Var}(\widehat{\mbox{CDE}}_{obs})=E[(\widehat{\mbox{CDE}}_{obs}-E[(\widehat{\mbox{CDE}}_{obs}])^2],\label{varcdehat}
\end{equation}
as cases of \textsl{statistical error} in that, unlike the other two, these depend only on the observed data distribution, sample size, and statistical model, irrespective of a causal question and causal model. 

\subsection{The inverse probability weighted estimator of Stensrud et al. }

The following alternative IP weighted estimator was posed by \citet{stensrud2022marginal} for a choice of $a_D\in\{0,1\}$:
\begin{align}
\widehat{\mbox{SDE}}^{a_D}_{obs} 
  &= \hat E\left.\left[Y\frac{\{1-\tilde{\pi}(A=a_D, L;\hat{\tilde{\beta}})\}}{\{1-\tilde{\pi}(A=1, L;\hat{\tilde{\beta}})\}}\right|A=1\right]-\hat E\left.\left[Y\frac{\{1-\tilde{\pi}(A=a_D, L;\hat{\tilde{\beta}})\}}{\{1-\tilde{\pi}(A=0, L;\hat{\tilde{\beta}})\}}\right|A=0\right]. \label{sdehat}
\end{align}

Given the premise that the actual causal target is \eqref{sdecontrast}, the total bias of this estimator is $\mbox{Bias}(\widehat{SDE}_{obs})_{SDE^{a_D}_0}=E[\widehat{\mbox{SDE}}^{a_D}_{obs}]- \mbox{SDE}^{a_D}_0$ with the decomposition: 
\begin{align}
\mbox{Bias}(\widehat{SDE}_{obs})_{SDE^{a_D}_0}&=\{\mbox{SDE}^{a_D}_{obs}-\mbox{SDE}^{a_D}_0\}\mbox{ } (\mbox{non-identification error})\nonumber\\
&\quad +\{E[\widehat{\mbox{SDE}}^{a_D}_{obs}]-\mbox{SDE}^{a_D}_{obs}\} \mbox{ } (\mbox{statistical error}).
  \label{biassdeobs}
\end{align}
In this case, estimand error is absent because the actual causal target is the ostensible causal target. Further, the non-identification error component of \eqref{biassdeobs} is distinct from that of \eqref{biascdeobs} pertaining to the IPCW estimator.  We will consider the difference in these two non-identification errors in more depth in Section \ref{section:non-identification}.  Analogously, we refer to the last component of \eqref{biassdeobs} as statistical error in the estimator \eqref{sdehat} along with its variance
\begin{equation}\mbox{Var}(\widehat{\mbox{SDE}}^{a_D}_{obs})=E[(\widehat{\mbox{SDE}}^{a_D}_{obs}-E[(\widehat{\mbox{SDE}}^{a_D}_{obs}])^2],\label{varsdehat}
\end{equation}
which is notably distinct from the variance measure \eqref{varcdehat}.  

\section{Illustrating the relative implications of the three error sources}

In this section, we illustrate relative implications of these three error sources for the two IP weighted estimators. Letting  $\mu_0(a,l,u)= \Pr(Y=1|A=a, L=l, U=u, D=0)$ and $\pi_0(a,l,u) = \Pr(D=1|A=a, L=l, U=u)$, we consider the following parametrization of the data generating distribution
\begin{align}
\mu_0(a,l,u; \theta)&= \mathrm{expit}(\theta_0 + \theta_1 a + \theta_2 l + \theta_3 a l + \theta_4 u + \theta_5 a u + \theta_6 lu) \label{ymodel} \\
\pi_0(a,l,u; \beta)&=\mathrm{expit}(\beta_0 + \beta_1 a + \beta_2 l + \beta_3 a l + \beta_4 u + \beta_5 a u + \beta_6 lu). \label{dmodel}
\end{align}
We quantify error sources for a range of scenarios distinguished by parameter value choices under this parametrization. In all scenarios, $L$ and $U$ are binary, with $\Pr(L=1)=0.5$.

\subsection{Implications of estimand error\label{section:identity slippage}}
\begin{figure}
\includegraphics[width=\textwidth, keepaspectratio]{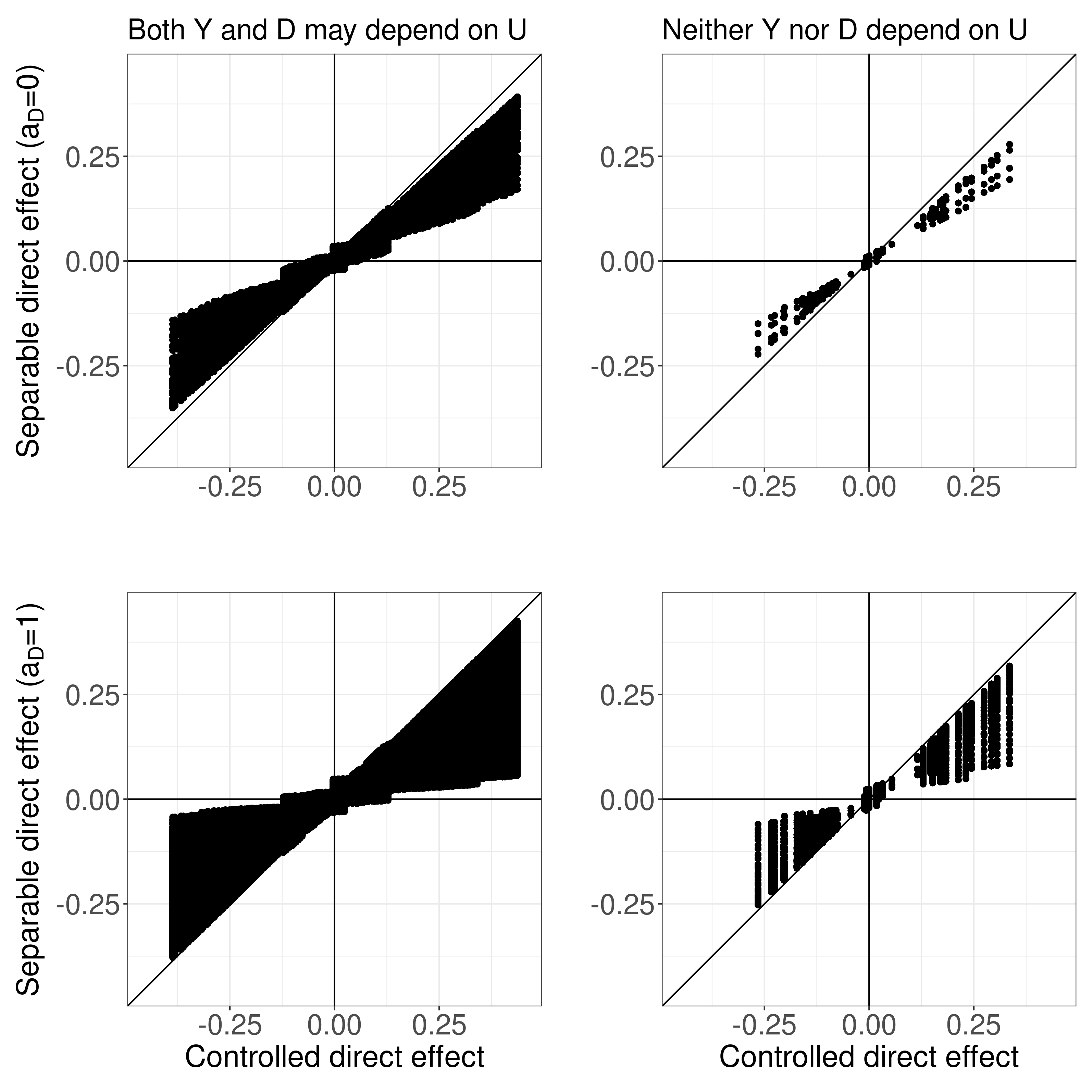}
\caption{Illustration of estimand error when the actual causal target is a separable direct in the scenarios where $D$ is non-rare (i.e., under the parameters given in Web Table 1 where $\Pr(D=1|A=a,L=l,U=u)\geq 10\%$ for some $(a,l,u)$).}
\label{fig:IdentitySlippage_nonrare}
\end{figure}

 Each panel in Figure \ref{fig:IdentitySlippage_nonrare} plots, for different combined values of the data generating parameters, the true value of the controlled direct effect \eqref{cdecontrast} against the true value of the separable direct effect \eqref{sdecontrast}, with their difference determining the magnitude of estimand error in \eqref{biascdeobs}. We set $\theta_i, \beta_i \in \{-1, -0.5, 0.5, 1\}$ for $i = 1, \dots, 6$ with further details in the Web Table 1. The simplicity of our setting allows calculating \eqref{cdecontrast} and \eqref{sdecontrast} analytically for any choice of data generating parameters under the models \eqref{ymodel} and \eqref{dmodel} (Web Appendix B).

Panels labeled ``Both $Y$ and $D$ may depend on $U$'' (allowing scenarios with non-identification error) versus ``Neither $Y$ nor $D$ depend on $U$''  (where non-identification error is zero) distinguish scenarios where $\Pr(U=1)=0.5$ versus $\Pr(U=1) =0$, respectively.  We further distinguish scenarios where the competing event is rare versus not rare controlled by the choice of $\beta_0$, with larger negative values aligned with a rare setting (see Web Table 1).

Points deviating from the 45\degree\,line indicate scenarios where the estimand error is non-zero. When the competing event is non-rare, we see substantial discrepancies between the ostensible and actual causal target can exist regardless of non-identification error. Points falling in the top left and bottom right quadrants of each panel reflect scenarios in which the ostensible and actual causal targets take different signs.  Such points do occur, albeit close to the origin, regardless of the dependence structure.  Not surprisingly, when competing events are rare, there is negligible estimand error (Web Figure 1).

\subsection{Relative implications of non-identification error \label{section:non-identification} } 
\begin{figure}
\includegraphics[width=\textwidth, keepaspectratio]{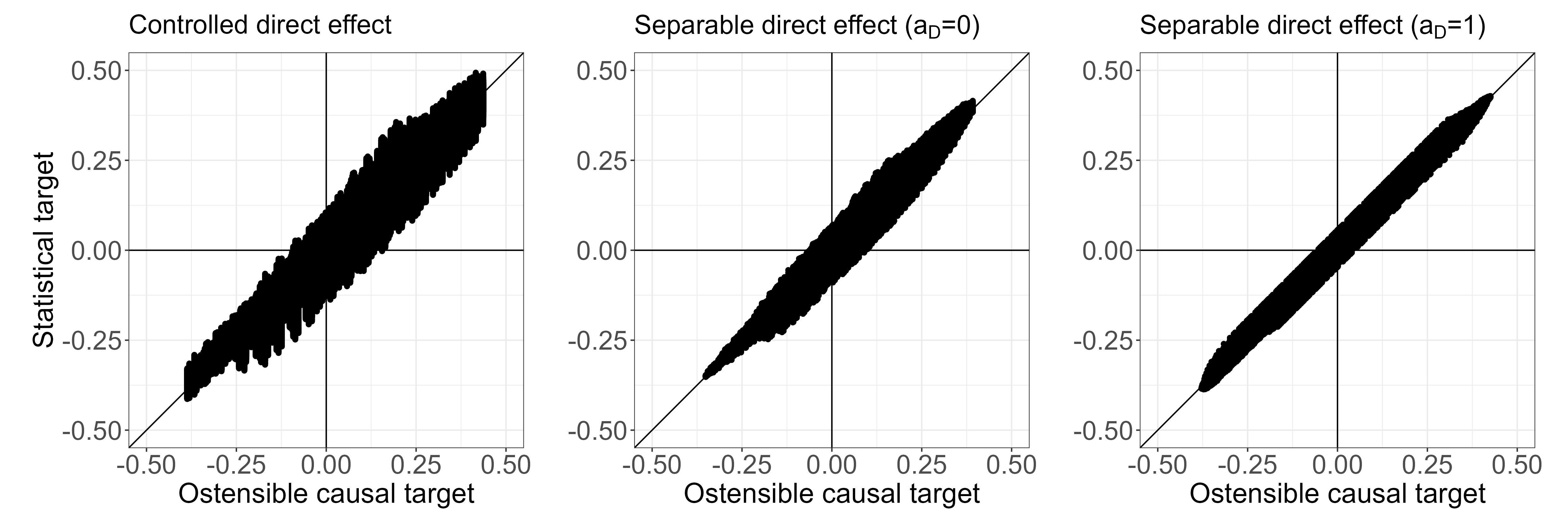}
\caption{Illustration of non-identification error as defined in \eqref{biascdeobs} where the ostensible causal target is a controlled direct effect (left) and non-identification error as defined in \eqref{biassdeobs} where the ostensible causal target is a separable direct effect for $a_D=0$ (center) and for $a_D=1$ (right).}
\label{fig:NIerror}
\end{figure}

For a range of parameters in our data generating mechanism, Figure \ref{fig:NIerror} plots the value of the ostensible causal target (x-axis) versus its corresponding statistical target (y-axis). Points were generated for each of the panels from the same combinations of parameter values such that all plots quantify non-identification error under the same range of data generating scenarios. Details are summarized in Web Table 1.  For all scenarios $\Pr(U=1)=0.5$. Again, for our simple setting, the magnitude of each point in Figure \ref{fig:NIerror} and, in turn, non-identification error was calculated analytically based on algebraic expressions derived from the true data-generating models (see Web Appendix B).

Larger deviations from the 45\degree\,line indicate larger non-identification error.  For the scenarios considered, we can see that the existence of an unmeasured shared cause of $Y$ and $D$ does not affect the non-identification error of a controlled and separable direct effect equally, with more extreme values of non-identification error in an analysis that censors competing events (i.e. where the ostensible target is a controlled direct effect) compared to an analysis designed for a separable direct effect.

\subsection{Relative implications on variance of near positivity violations: a simulation study  \label{simulation}}
Following arguments in Section \ref{causalstory}, identification of a controlled direct effect by \eqref{cdeobscontrast} and a separable direct effect by \eqref{sdeobscontrast} rely on different positivity conditions.  In particular, in a setting where \eqref{pos} fails, the controlled direct effect will not be identified while a separable direct effect may still be identified.  Even in settings where \eqref{pos} theoretically holds, so-called \textsl{near} violations of this condition \citep{petersenpos} will correspondingly have a greater impact on the variance \eqref{varcdehat} compared to the variance \eqref{varsdehat}.  Here, we will say that such a near violation exists when, for some $A=a,L=l$, $\{1- \tilde{\pi}(a, l)\}$ is positive yet close to zero.  

To investigate the difference in this source of statistical error between the estimators \eqref{cdehat} and  \eqref{sdehat}, we conducted a simulation study. The simulations were based on $20,000$ samples of $n=100,000$ independent and identically distributed observations.  Data on $(U,L,A,D,Y)$ for each observation was generated as follows: $A$ and $U$ were independently drawn from Bernoulli(0.5); $L$ was drawn from Bernoulli(0.1). $D$ was drawn from a logistic model $\pi_0(a,l,u; \beta)$ \eqref{dmodel}, using specified coefficients $\beta$. If $D=0$, $Y$ was drawn from a logistic model $\mu_0(a,l,u; \theta)$ \eqref{ymodel}, using specified coefficients $\theta$; otherwise, if $D=1$, we set $Y=0$. To focus on the variance component of statistical error, we used a common saturated model for $\tilde{\pi}(A, L;\tilde{\beta})$ in calculating each of the two IP weighted estimators \eqref{cdehat} and \eqref{sdehat} to ensure correct specification of the observed data nuisance function \eqref{psdeath}. 

We simulated data in the presence versus the absence of near positivity violations as well as scenarios with variation in the dependence of $Y$ and $D$ on the unmeasured $U$ and \textsl{marginally} rare ($\Pr(D=1)<10\%$) versus \textsl{marginally} non-rare competing events ($\Pr(D=1)\geq 10\%$).  The specific coefficient values used in each of the scenarios are provided in Web Table 2.  In Table \ref{table:simulation_results_stderr}, we present a comparison of the variance of the IPCW estimator and the variance of the alternative weighted estimator for $a_D=0$ (results for $a_D=1$ are provided in Web Table 3). 

\begin{table}[ht]
\caption{Simulation-based comparison of the variance Var($\widehat{\mbox{SDE}}^{a_D=0}_{obs}$) versus  Var($\widehat{\mbox{CDE}}_{obs})$. Expectations in the variance calculations were taken relative to the distribution over simulation runs.}
\centering
\begin{tabular}{ccccc}
  \hline
\begin{tabular}{c} Near \\ positivity \\ violation? \end{tabular}  & 
\begin{tabular}{c} Dependence of\\ $Y$ and $D$\\ on $U$? \end{tabular} & 
\begin{tabular}{c} Competing\\ events \\ marginally rare? \end{tabular} & 
Var($\widehat{\mbox{SDE}}^{a_D=0}_{obs}$) &
Var($\widehat{\mbox{CDE}}_{obs})$
\\
  \hline
No & No & Yes &  $6 \times10^{-6}$ & $7 \times10^{-6}$ \\ 
  No & No & No & $4 \times10^{-6}$ & $9 \times10^{-6}$ \\ 
  No & Yes & No & $5 \times10^{-6}$ & $1 \times10^{-5}$ \\ 
  Yes & No & Yes & $4 \times10^{-6}$ & $4 \times10^{-4}$ \\ 
  Yes & No & No & $4 \times10^{-6}$ & $4 \times10^{-4}$ \\ 
  Yes & Yes & No & $4 \times10^{-6}$ & $1 \times10^{-4}$ \\ 
   \hline
\end{tabular}
\label{table:simulation_results_stderr}
\end{table}

We can see from Table \ref{table:simulation_results_stderr} that the variance of the IPCW estimator \eqref{cdehat} is larger than the variance of the alternative estimator \eqref{sdehat} in all scenarios.  As expected, the difference in the variance of the two estimators is especially pronounced in the presence of near positivity violations with the variance \eqref{varcdehat} of the IPCW estimator up to 100 times that of the variance of the alternative weighted estimator derived with a separable direct effect in mind, regardless of whether competing events are marginally rare or non-rare. 

\section{Comparative empirical analysis of the real randomized trial of estrogen therapy  \label{analysis}}
\citet{young2020causal} applied the IPCW estimator \eqref{cdehat}, and \citet{stensrud2022marginal} the alternative weighted estimator \eqref{sdehat}, both generalized to accommodate the true time-varying nature of competing event and event of interest processes, to data from the real randomized trial of estrogen therapy and prostate cancer death \citep{prostatedata}. In this section, we present a direct empirical comparison of these two approaches in light of the theoretical comparison of bias sources discussed above. 

We refer the reader to Web Appendix E for details.  Briefly, as in both \citet{young2020causal} and \citet{stensrud2022marginal}, our analysis was restricted to the 125 patients randomized to the high-dose DES arm ($A=1$) and the 127 patients randomized to placebo ($A=0$). Moreover, the competing event hazard at each time conditional on $A$ and $L$ (generalizing the nuisance function \eqref{psdeath} for the true time-varying data structure) was assumed to follow a pooled logistic model dependent on time (as a second-degree polynomial function of month), treatment $A$ and covariates $L$,  including dichotomized serum hemoglobin ($<$12  vs. $\geq$12, g/100ml), indicators of age group ($\leq$59, 60 to 75, $\geq$75, years old), activity level (normal activity vs. in bed), and history of cardiovascular disease.  Our analysis differs slightly from those presented in \citet{young2020causal} and \citet{stensrud2022marginal} in that we restricted the follow-up period to the first 50 months based on the fact that no individuals were lost to follow-up during that time, thereby avoiding censoring due to loss to follow-up. Additionally, we included an interaction term between $A$ and history of cardiovascular disease (a component of $L$ in the pooled logistic model), as the presence of such interaction terms can increase divergence between the identifying functionals of the controlled direct and separable direct effects  (see Web Appendix F).

\begin{figure}
\includegraphics[width=\textwidth, keepaspectratio]{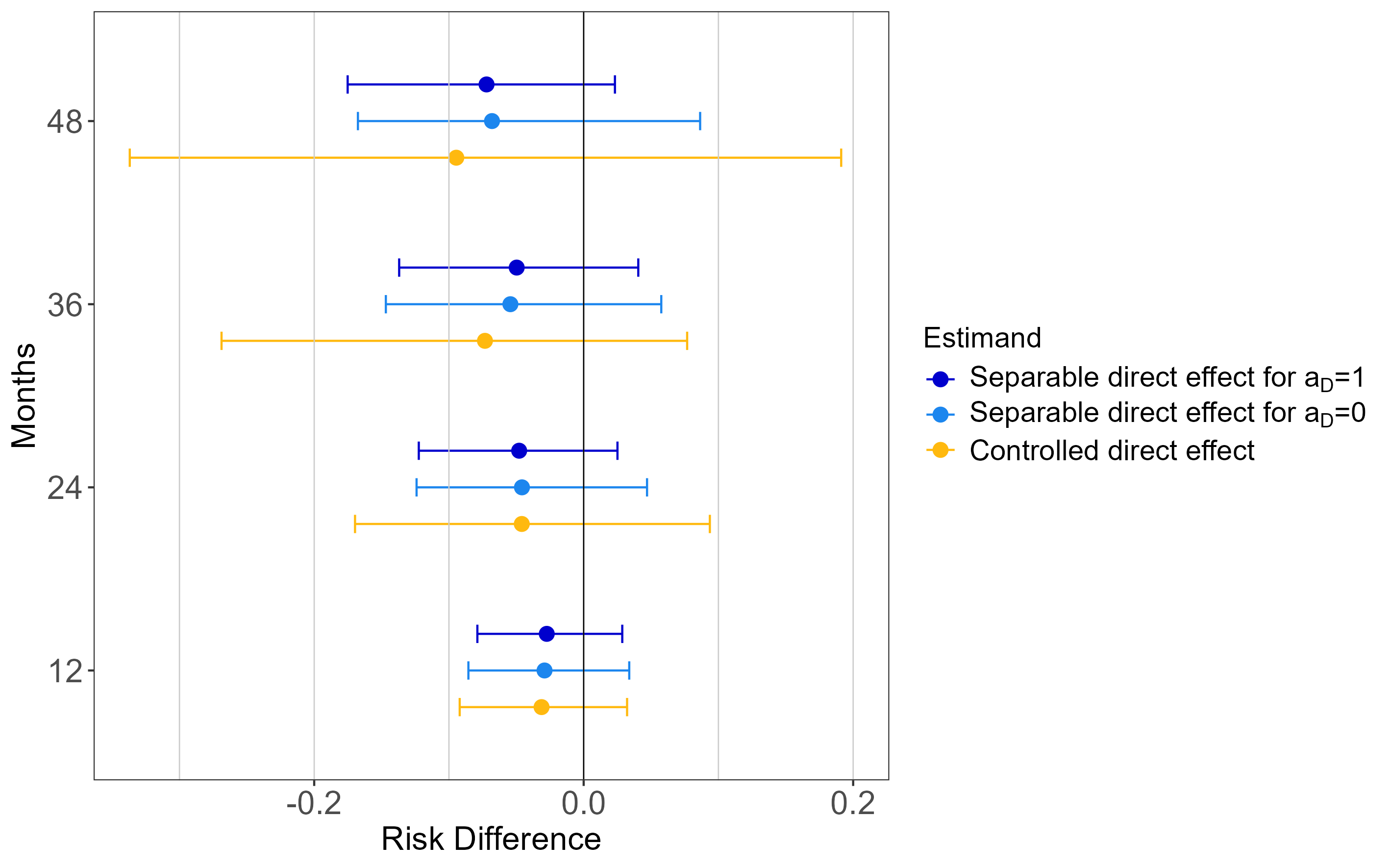}
\caption{The separable and controlled direct effects from the randomized trial of estrogen therapy dataset at 12, 24, 36 and 48 months. The 95\% confidence intervals are obtained from 1000 bootstrap samples by taking the 2.5th and 97.5th percentiles of the estimates.}
\label{fig:prostate-forest}
\end{figure}

Figure \ref{fig:prostate-forest} presents point estimates and 95\% confidence intervals for the ostensible causal targets of the IPCW estimator (a controlled direct effect) and the alternative weighted estimator of \citet{stensrud2022marginal} (a separable direct effect, for $a_D=1$ and $a_D=0$) at 12, 24, 36, and 48 months. The confidence intervals are obtained from 1000 bootstrap samples by taking the 2.5th and 97.5th percentiles of the estimates. Point estimates are relatively similar up to 24 months after which we see that controlled direct effect estimates become more extreme than the separable direct effect estimates. 

The maximum weight for the IPCW estimator in our analysis (14.4) was much larger than the maximum weight for the separable direct effect estimator (3.4). Correspondingly and in line with our simulation findings, the 95\% confidence intervals for the IPCW estimates of the controlled direct effect were wider even at follow-up times where the point estimates were similar (12 and 24 months).

\section{Discussion}\label{discussion}

In settings with competing events, investigators may be interested in a direct treatment effect on an event of interest that does not capture treatment effects via competing events. Investigators should choose a notion of a direct treatment effect based on their underlying scientific question. In this paper, we have concretely illustrated the cascading consequences of conflating an effect referencing a scenario where somehow competing events are universally eliminated, with a separable direct effect, an effect of a modified version of the current study treatment.  The importance of making the distinction between these two effects apparent is reinforced by recent claims of their equivalence \citep{austin2025inverse}. Our presentation clarifies that these two effects answer substantively different questions, rely on different identifying assumptions, and their corresponding estimators have different properties. 

We distinguished between two distinct sources of causal bias in an analysis: interpretational error versus non-identification error.  While considerations of bias due to causal non-identification have a long history in the causal inference and statistics literature, interpretational error has received relatively little attention.  A notable exception is the recent formalization of the notion of \textsl{identity slippage} \citep{sarvet2023interpretational}. Identity slippage might be understood as a special case of estimand error as we have defined it here, specifically, the special case where the actual and ostensible causal targets differ in the presence of non-identification error but coincide when both targets are identified.  Our case of estimand error might be characterized as even more extreme in the sense that the controlled direct effect and separable direct effect are not generally equal even in settings where both are identified (that is, where non-identification error is zero).  Interpretational error such as this is avoidable in any study by engaging up front with an investigator's, often implicit, motivating causal story prior to, and separate from, the data available for analysis and the specifics of how that data will be analyzed \citep{young2024story}.  

Non-identification error to some degree is generally unavoidable, particularly when the underlying causal story involves treatment mechanism, but also more broadly.  Understanding the actual causal target prior to study design and data collection is the best way to minimize this form of error.  We have clarified that not all non-identification error is ``created equal''.  Under the range of data generating mechanisms we considered, the controlled direct effect was generally subject to larger non-identification error than the separable direct effect due to the presence of an unmeasured shared cause of the competing event and event of interest.   This finding cannot be clearly generalized to all settings.  Still, there are some possible explanations for this observation as a more general phenomenon.  First, in a randomized trial where $A$ is randomized and there is perfect adherence to the randomization (i.e. ``$A$ is unconfounded''), both components of the counterfactual mean difference that comprise the controlled direct effect are unidentified when $U$ in Figure 1 is present.  By contrast, for a separable direct effect indexed by a choice of $a_D$, the component of the counterfactual mean difference that comprises this separable direct effect with $a_Y=a_D$ will remain identified even when $U$ is a shared cause of $Y$ and $D$. It is only the component of the difference referencing $a_Y\neq a_D$ that compromises identification when $U$ is a shared cause. Second, the lack of inclusion of $U$ in estimation of the nuisance function \eqref{psdeath} may generally have a greater impact on the IP of censoring weighted estimator \eqref{cdehat}, where an estimate of \eqref{psdeath} only contributes to the denominator, compared to the alternative weighted estimator \eqref{sdehat}, where this estimate is required in both the numerator and denominator, possibly leading to some degree of bias cancellation.  Future work might consider further exploration of this finding.

We limited our presentation to consideration of error in singly robust weighted estimators of a controlled direct effect and separable direct effect due to their computational and structural simplicity. Fully parametric g-computation methods \citep{mcgrath2020gformula,stensrud2021generalized,stensrud2022marginal}, as well as more robust estimators based on the efficient influence function \citep{unified,van2018targeted,chernozhukov2018double} are also available. The choice of estimator has no bearing on our formalization and illustration of the magnitude of either ``causal'' component of bias (interpretational or non-identification error).  We conjecture that our conclusions regarding greater impacts on variance due to near positivity violations for the singly robust IP of censoring weighted estimator extend to any estimator that ``censors'' competing events as opposed to a comparable estimator for targeting a separable direct effect.  This is based on the fact that the separable direct effect and controlled direct effect identifying functionals coincide with cases of a g-formula for a stochastic versus a deterministic treatment intervention, respectively, with competing events in the ``role'' of treatment. There is a growing literature emphasizing and illustrating the performance benefits of targeting stochastic g-formula functionals in settings where targeting deterministic functionals breaks down \citep{young2014identification,kennedy2019nonparametric,diaz2023nonparametric}.

Finally, we illustrated the cascading implications of misaligning questions and statistics under a particular premise on the actual causal target such that true interest is in the separable effect \eqref{sdecontrast} under mechanistic assumptions on the modified treatments $A_Y$ and $A_D$ under full isolation that give this contrast a ``direct'' effect interpretation. Full isolation is importantly not necessary for identifying \eqref{sdecontrast} and weaker isolation conditions are permitted \citep{stensrud2021generalized}.  In some settings, separable effects under weaker isolation conditions will best align with what motivates an investigator.  In others, the actual causal target might be something completely different than a separable effect.  Regardless, controlled direct effects will generally not align with the actual causal target when universal elimination of competing events is implausible. In conclusion, whatever the actual causal target, failing to make the actual causal target explicit prior to decisions on data analysis inevitably leads to avoidable estimand error and may amplify other unavoidable errors.


\section*{Acknowledgements}

We thank Dr. Mats Stensrud for helpful discussions and various comments. T. Kawahara is funded by the UTokyo Global Activity Support Program for Young Researchers and partially by JSPS KAKENHI Grant Number 22K17301. T. Kawahara was supported by Thomas O. Pyle Fellowship from the Harvard Pilgrim Health Care Institute. \vspace*{-8pt}

\bibliographystyle{biom} 
\bibliography{refs}

\renewcommand{\tablename}{Web Table}
\renewcommand{\figurename}{Web Figure}
\renewcommand{\theequation}{A\arabic{equation}}
\setcounter{table}{0}
\setcounter{figure}{0}
\setcounter{equation}{0}

\newpage
\begin{center}
{\Large \textbf{Supplementary Materials for ``Illustrating implications of misaligned causal questions and statistics in settings with competing events and interest in treatment mechanisms'' by Takuya Kawahara, Sean McGrath and Jessica G. Young}}
\end{center}

\section*{Web Appendix A. Data availability}
The data and code used in this work are publicly available at: \newline https://github.com/TakuyaKawahara/Misaligned-questions-and-statistics-paper.

\section*{Web Appendix B. Analytical formulas for causal source of error \label{formulas}}

Let us return to the causal model communicated implicitly by Figure 1 and explicitly by Figures 2b-d where, for simplicity, we assume both $L$ and $U$ are binary variables.  Denoting 
\begin{align*}
\mu_0(a,l,u) &= \Pr(Y=1|A=a, L=l, U=u, D=0) \\
\pi_0(a,l,u) &= \Pr(D=1|A=a, L=l, U=u) \\
p_L &= \Pr(L=1) \\
p_U &= \Pr(U=1) 
\end{align*} we can alternatively write (2) as
\begin{align}
  \psi_0(a,d=0)&= \mu_0(a,1,1)p_Lp_U + \mu_0(a,1,0)p_L(1-p_U) \notag\\ 
  &\quad +\mu_0(a,0,1)(1-p_L)p_U + \mu_0(a,0,0)(1-p_L)(1-p_U)\label{cdefactual_example}
  \end{align}
  and (8) as
  \begin{align}
  \psi_0(a_Y,a_D)&= \mu_0(a_Y,1,1)\{1-\pi_0(a_D,1,1)\}p_Lp_U  \notag \\
  &\quad + \mu_0(a_Y,1,0)\{1-\pi_0(a_D,1,0)\}p_L(1-p_U) \notag \\
  &\quad + \mu_0(a_Y,0,1)\{1-\pi_0(a_D,0,1)\}(1-p_L)p_U  \notag \\
  &\quad + \mu_0(a_Y,0,0)\{1-\pi_0(a_D,1,0)\}(1-p_L)(1-p_U)\label{sdefactual_example}
\end{align}
respectively. 

By plugging \eqref{cdefactual_example} and \eqref{sdefactual_example} into the definition of \textit{estimand error},
\begin{align}
\mbox{CDE}_0 - \mbox{SDE}^{a_D}_0 &=\{\psi_0(a=1,d=0)-\psi_0(a=0,d=0)\} \notag\\
 &\quad -\{\psi_0(a_Y=1,a_D)-\psi_0(a_Y=0,a_D)\} \label{cde-estimand-error},
\end{align}
we obtain an expression in terms of the joint distribution of $(U,L,A,D,Y)$,
\begin{align}
    \mbox{CDE}_0 - \mbox{SDE}^{a_D}_0 = &\{\mu_0(1,1,1) - \mu_0(0,1,1) \} \pi_0 (a_D,1,1)p_Lp_U \notag\\
    & + \{\mu_0(1,1,0) - \mu_0(0,1,0) \} \pi_0 (a_D,1,0)p_L(1-p_U) \notag\\
    & + \{\mu_0(1,0,1) - \mu_0(0,0,1) \} \pi_0 (a_D,0,1)(1-p_L)p_U \notag\\    
    & + \{\mu_0(1,0,0) - \mu_0(0,0,0) \} \pi_0 (a_D,0,0)(1-p_L)(1-p_U) 
\label{EstimandError_cde0sde0}
\end{align}

Similarly, we obtain an expression for the \textit{non-identification error} in any estimator for the controlled direct effect,
\begin{align}
\{\mbox{CDE}_{obs} - \mbox{CDE}_0\} &=\{\tilde\psi(a=1,d=0)-\tilde\psi(a=0,d=0)\}  \notag\\ 
        &\quad -\{\psi_0(a=1,d=0)-\psi_0(a=0,d=0)\},    \label{cde-non-identification error}
\end{align}
as
\begin{align}
   &\{\mbox{CDE}_{obs}-\mbox{CDE}_0\} \notag\\
    =  & -\dfrac{\pi_0 (1,1,1)-\pi_0 (1,1,0)}{\{1-\pi_0 (1,1,1)\}p_U+\{1-\pi_0 (1,1,0)\}(1-p_U)}\left\{\mu_0 (1,1,1) - \mu_0 (1,1,0)\right\}p_Lp_U(1-p_U) \nonumber\\
    & -\dfrac{\pi_0 (1,0,1)-\pi_0 (1,0,0)}{\{1-\pi_0 (1,0,1)\}p_U+\{1-\pi_0 (1,0,0)\}(1-p_U)}\left\{\mu_0 (1,0,1) - \mu_0 (1,0,0)\right\}(1-p_L)p_U(1-p_U) \nonumber\\
    & +\dfrac{\pi_0 (0,1,1)-\pi_0 (0,1,0)}{\{1-\pi_0 (0,1,1)\}p_U+\{1-\pi_0 (0,1,0)\}(1-p_U)}\left\{\mu_0 (0,1,1) - \mu_0 (0,1,0)\right\}p_Lp_U(1-p_U) \nonumber\\
    & +\dfrac{\pi_0 (0,0,1)-\pi_0 (0,0,0)}{\{1-\pi_0 (0,0,1)\}p_U+\{1-\pi_0 (0,0,0)\}(1-p_U)}\left\{\mu_0 (0,0,1) - \mu_0 (0,0,0)\right\}(1-p_L)p_U(1-p_U)
\label{NI_cde}
\end{align}
and for the non-identification error \eqref{sde-non-identification error} in any estimator for the separable direct effects,
\begin{align}
    \{\mbox{SDE}^{a_D}_{obs} - \mbox{SDE}^{a_D}_0\}&=\{\tilde\psi(a_Y=1,a_D)-\tilde\psi(a_Y=0,a_D)\}  \notag\\ 
        &\quad -\{\psi_0(a_Y=1,a_D)-\psi_0(a_Y=0,a_D), \label{sde-non-identification error}
\end{align}
as
\begin{align}
 &\{\mbox{SDE}^{a_D}_{obs}-\mbox{SDE}^{a_D}_0\} \notag\\
   & = \dfrac{\{1-\pi_0 (1,1,1)\}\{1-\pi_0 (a_D,1,0)\}-\{1-\pi_0 (a_D,1,1)\}\{1-\pi_0 (1,1,0)\}}{\{1-\pi_0 (1,1,1)\}p_U + \{1-\pi_0 (1,1,0)\}(1-p_U)}\nonumber\\
    &\quad\quad\quad \times\left\{\mu_0 (1,1,1) - \mu_0 (1,1,0)\right\}p_Lp_U(1-p_U) \nonumber\\
&\quad+\dfrac{\{1-\pi_0 (1,0,1)\}\{1-\pi_0 (a_D,0,0)\}-\{1-\pi_0 (a_D,0,1)\}\{1-\pi_0 (1,0,0)\}}{\{1-\pi_0 (1,0,1)\}p_U + \{1-\pi_0 (1,0,0)\}(1-p_U)}\nonumber\\
    &\quad\quad\quad \times\left\{\mu_0 (1,0,1) - \mu_0 (1,0,0)\right\}(1-p_L)p_U(1-p_U) \nonumber\\
&\quad-\dfrac{\{1-\pi_0 (0,1,1)\}\{1-\pi_0 (a_D,1,0)\}-\{1-\pi_0 (a_D,1,1)\}\{1-\pi_0 (0,1,0)\}}{\{1-\pi_0 (0,1,1)\}p_U + \{1-\pi_0 (0,1,0)\}(1-p_U)}\nonumber\\
    &\quad\quad\quad \times\left\{\mu_0 (0,1,1) - \mu_0 (0,1,0)\right\}p_Lp_U(1-p_U) \nonumber\\
&\quad-\dfrac{\{1-\pi_0 (0,0,1)\}\{1-\pi_0 (a_D,0,0)\}-\{1-\pi_0 (a_D,0,1)\}\{1-\pi_0 (0,0,0)\}}{\{1-\pi_0 (0,0,1)\}p_U + \{1-\pi_0 (0,0,0)\}(1-p_U)}\nonumber\\
    &\quad\quad\quad \times\left\{\mu_0 (0,0,1) - \mu_0 (0,0,0)\right\}(1-p_L)p_U(1-p_U). 
\label{NI_sde}
\end{align}
See Web Appendix C for detailed derivations. Note that, in equation \eqref{NI_sde}, the first two terms in the sum disappear when we consider the separable direct effect evaluated at $a_D=1$, while the last two terms in the sum disappear when evaluated at $a_D=0$.

\section*{Web Appendix C. Derivation of \eqref{NI_cde} and \eqref{NI_sde} \label{derivation}}

\subsection*{Derivation of \eqref{NI_cde}}

By definition, the non-identification error is given by
\begin{align*}
    \mbox{CDE}_{obs}-\mbox{CDE}_0 & =\{\tilde\psi(a=1,d=0)-\tilde\psi(a=0,d=0)\}  \notag\\ 
        &\quad -\{\psi_0(a=1,d=0)-\psi_0(a=0,d=0)\}
\end{align*}
where 
\begin{align}
&\psi_0(a=1,d=0)-\psi_0(a=0,d=0)  \nonumber\\
&=\mu_0 (1,1,1)p_Lp_U + \mu_0 (1,1,0)p_L(1-p_U)  \nonumber\\
  &\quad+\mu_0 (1,0,1)(1-p_L)p_U +\mu_0 (1,0,0)(1-p_L)(1-p_U)  \nonumber\\
  &\quad-\mu_0 (0,1,1)p_Lp_U - \mu_0 (0,1,0)p_L(1-p_U)  \nonumber\\
  &\quad- \mu_0 (0,0,1)(1-p_L)p_U -\mu_0 (0,0,0)(1-p_L)(1-p_U)     \label{app_cde_correct}
\end{align}

We next derive a suitable expression for $\tilde\psi(a=1,d=0) - \tilde\psi(a=0,d=0)$, where recall that $\tilde\psi(a,d=0) = \sum_{l} \Pr(Y=1|D=0, A=a, L=l) \Pr(L=l)$. To do so, it will be convenient to express $\mu(a,l):=\Pr(Y=1|D=0, A=a, L=l)$ in terms of $\mu_0(a,l,u)$. By standard probability laws and the independence assumptions encoded in the DAG,
\begin{align}
\mu(a,l) &=\sum_{u}\Pr(Y=1|D=0, A=a, L=l, U=u) \Pr(U=u|D=0, A=a, L=l) \nonumber\\
 &=\sum_{u}\Pr(Y=1|D=0, A=a, L=l, U=u) \times \nonumber\\
  &\quad\quad \dfrac{\Pr(D=0|A=a, L=l, U=u)\Pr(L=l,U=u|A=a)}{\sum_{u}\Pr(D=0|A=a, L=l, U=u)\Pr(L=l,U=u|A=a)} \nonumber\\
 &=\dfrac{\sum_{u}\mu_0 (a,l,u)\{1-\pi_0 (a,l,u)\}\Pr(L=l,U=u)}{\sum_{u}\{1-\pi_0 (a,l,u)\}\Pr(L=l,U=u)} \nonumber \\
 &=\dfrac{\sum_{u}\mu_0 (a,l,u)\{1-\pi_0 (a,l,u)\}\Pr(U=u)}{\sum_{u}\{1-\pi_0 (a,l,u)\}\Pr(U=u)} \label{app_mu_withoutu}
\end{align}

Plugging \eqref{app_mu_withoutu} into the expression for $\tilde\psi(a,d=0)$, 
\begin{align*}
&\tilde\psi(a=1, d=0) - \tilde\psi(a=0, d=0) \notag\\
 &= \dfrac{\mu_0 (1,1,1)\{1-\pi_0 (1,1,1)\}p_U + \mu_0 (1,1,0)\{1-\pi_0 (1,1,0)\}(1-p_U)}{\{1-\pi_0 (1,1,1)\}p_U + \{1-\pi_0 (1,1,0)\}(1-p_U)}p_L  \\
  &\quad + \dfrac{\mu_0 (1,0,1)\{1-\pi_0 (1,0,1)\}p_U + \mu_0 (1,0,0)\{1-\pi_0 (1,0,0)\}(1-p_U)}{\{1-\pi_0 (1,0,1)\}p_U + \{1-\pi_0 (1,0,0)\}(1-p_U)}(1-p_L)  \\
  &\quad - \dfrac{\mu_0 (0,1,1)\{1-\pi_0 (0,1,1)\}p_U + \mu_0 (0,1,0)\{1-\pi_0 (0,1,0)\}(1-p_U)}{\{1-\pi_0 (0,1,1)\}p_U + \{1-\pi_0 (0,1,0)\}(1-p_U)}p_L \\
  &\quad - \dfrac{\mu_0 (0,0,1)\{1-\pi_0 (0,0,1)\}p_U + \mu_0 (0,0,0)\{1-\pi_0 (0,0,0)\}(1-p_U)}{\{1-\pi_0 (0,0,1)\}p_U + \{1-\pi_0 (0,0,0)\}(1-p_U)}(1-p_L) 
\end{align*}
which can be re-expressed as
\begin{align}
& \tilde\psi(a=1, d=0) - \tilde\psi(a=0, d=0) \notag\\
  &=\dfrac{1-\pi_0 (1,1,1)}{\{1-\pi_0 (1,1,1)\}p_U + \{1-\pi_0 (1,1,0)\}(1-p_U)}\mu_0 (1,1,1)p_Lp_U \nonumber\\
  &\quad+\dfrac{1-\pi_0 (1,1,0)}{\{1-\pi_0 (1,1,1)\}p_U + \{1-\pi_0 (1,1,0)\}(1-p_U)}\mu_0 (1,1,0)p_L(1-p_U)  \nonumber\\
  &\quad+\dfrac{1-\pi_0 (1,0,1)}{\{1-\pi_0 (1,0,1)\}p_U + \{1-\pi_0 (1,0,0)\}(1-p_U)}\mu_0 (1,0,1)(1-p_L)p_U \nonumber\\
  &\quad+\dfrac{1-\pi_0 (1,0,0)}{\{1-\pi_0 (1,0,1)\}p_U + \{1-\pi_0 (1,0,0)\}(1-p_U)}\mu_0 (1,0,0)(1-p_L)(1-p_U)  \nonumber\\
  &\quad -\dfrac{1-\pi_0 (0,1,1)}{\{1-\pi_0 (0,1,1)\}p_U + \{1-\pi_0 (0,1,0)\}(1-p_U)}\mu_0 (0,1,1)p_Lp_U \nonumber\\
  &\quad-\dfrac{1-\pi_0 (0,1,0)}{\{1-\pi_0 (0,1,1)\}p_U + \{1-\pi_0 (0,1,0)\}(1-p_U)}\mu_0 (0,1,0)p_L(1-p_U)  \nonumber\\
  &\quad-\dfrac{1-\pi_0 (0,0,1)}{\{1-\pi_0 (0,0,1)\}p_U + \{1-\pi_0 (0,0,0)\}(1-p_U)}\mu_0 (0,0,1)(1-p_L)p_U \nonumber\\
  &\quad-\dfrac{1-\pi_0 (0,0,0)}{\{1-\pi_0 (0,0,1)\}p_U + \{1-\pi_0 (0,0,0)\}(1-p_U)}\mu_0 (0,0,0)(1-p_L)(1-p_U)     \label{app_cde_incorrect}
\end{align}

Plugging  \eqref{app_cde_correct} and \eqref{app_cde_incorrect} into the expression for $\{\mbox{CDE}_{obs}-\mbox{CDE}_0\}$, 
\begin{align*}
&\{\mbox{CDE}_{obs}-\mbox{CDE}_0\}\notag\\
&=\left[\dfrac{1-\pi_0 (1,1,1)}{\{1-\pi_0 (1,1,1)\}p_U + \{1-\pi_0 (1,1,0)\}(1-p_U)}-1\right]\mu_0 (1,1,1)p_Lp_U \nonumber\\
  &\quad+\left[\dfrac{1-\pi_0 (1,1,0)}{\{1-\pi_0 (1,1,1)\}p_U + \{1-\pi_0 (1,1,0)\}(1-p_U)}-1\right]\mu_0 (1,1,0)p_L(1-p_U)  \nonumber\\
  &\quad+\left[\dfrac{1-\pi_0 (1,0,1)}{\{1-\pi_0 (1,0,1)\}p_U + \{1-\pi_0 (1,0,0)\}(1-p_U)}-1\right]\mu_0 (1,0,1)(1-p_L)p_U \nonumber\\
  &\quad+\left[\dfrac{1-\pi_0 (1,0,0)}{\{1-\pi_0 (1,0,1)\}p_U + \{1-\pi_0 (1,0,0)\}(1-p_U)}-1\right]\mu_0 (1,0,0)(1-p_L)(1-p_U)  \nonumber\\
  &\quad -\left[\dfrac{1-\pi_0 (0,1,1)}{\{1-\pi_0 (0,1,1)\}p_U + \{1-\pi_0 (0,1,0)\}(1-p_U)}-1\right]\mu_0 (0,1,1)p_Lp_U \nonumber\\
  &\quad-\left[\dfrac{1-\pi_0 (0,1,0)}{\{1-\pi_0 (0,1,1)\}p_U + \{1-\pi_0 (0,1,0)\}(1-p_U)}-1\right]\mu_0 (0,1,0)p_L(1-p_U)  \nonumber\\
  &\quad-\left[\dfrac{1-\pi_0 (0,0,1)}{\{1-\pi_0 (0,0,1)\}p_U + \{1-\pi_0 (0,0,0)\}(1-p_U)}-1\right]\mu_0 (0,0,1)(1-p_L)p_U \nonumber\\
  &\quad-\left[\dfrac{1-\pi_0 (0,0,0)}{\{1-\pi_0 (0,0,1)\}p_U + \{1-\pi_0 (0,0,0)\}(1-p_U)}-1\right]\mu_0 (0,0,0)(1-p_L)(1-p_U) \nonumber   
\end{align*}
Upon algebraic simplification,

\begin{align*}
&\{\mbox{CDE}_{obs}-\mbox{CDE}_0\}\notag\\
&=\dfrac{-\{\pi_0 (1,1,1)-\pi_0(1,1,0)\}(1-p_U)}{\{1-\pi_0 (1,1,1)\}p_U + \{1-\pi_0 (1,1,0)\}(1-p_U)}\mu_0 (1,1,1)p_Lp_U \nonumber\\
  &\quad+\dfrac{\{\pi_0 (1,1,1)-\pi_0(1,1,0)\}p_U}{\{1-\pi_0 (1,1,1)\}p_U + \{1-\pi_0 (1,1,0)\}(1-p_U)}\mu_0 (1,1,0)p_L(1-p_U)  \nonumber\\
  &\quad+\dfrac{-\{\pi_0 (1,0,1)-\pi_0(1,0,0)\}(1-p_U)}{\{1-\pi_0 (1,0,1)\}p_U + \{1-\pi_0 (1,0,0)\}(1-p_U)}\mu_0 (1,0,1)(1-p_L)p_U \nonumber\\
  &\quad+\dfrac{\{\pi_0 (1,0,1)-\pi_0(1,0,0)\}p_U}{\{1-\pi_0 (1,0,1)\}p_U + \{1-\pi_0 (1,0,0)\}(1-p_U)}\mu_0 (1,0,0)(1-p_L)(1-p_U)  \nonumber\\
  &\quad -\dfrac{-\{\pi_0 (0,1,1)-\pi_0(0,1,0)\}(1-p_U)}{\{1-\pi_0 (0,1,1)\}p_U + \{1-\pi_0 (0,1,0)\}(1-p_U)}\mu_0 (0,1,1)p_Lp_U \nonumber\\
  &\quad-\dfrac{\{\pi_0 (0,1,1)-\pi_0(0,1,0)\}p_U}{\{1-\pi_0 (0,1,1)\}p_U + \{1-\pi_0 (0,1,0)\}(1-p_U)}\mu_0 (0,1,0)p_L(1-p_U)  \nonumber\\
  &\quad-\dfrac{-\{\pi_0 (0,0,1)-\pi_0(0,0,0)\}(1-p_U)}{\{1-\pi_0 (0,0,1)\}p_U + \{1-\pi_0 (0,0,0)\}(1-p_U)}\mu_0 (0,0,1)(1-p_L)p_U \nonumber\\
  &\quad-\dfrac{\{\pi_0 (0,0,1)-\pi_0(0,0,0)\}p_U}{\{1-\pi_0 (0,0,1)\}p_U + \{1-\pi_0 (0,0,0)\}(1-p_U)}\mu_0 (0,0,0)(1-p_L)(1-p_U) \nonumber   \\  
&= -\dfrac{\pi_0 (1,1,1)-\pi_0 (1,1,0)}{\{1-\pi_0 (1,1,1)\}p_U+\{1-\pi_0 (1,1,0)\}(1-p_U)}\{\mu_0 (1,1,1)-\mu_0 (1,1,0)\}p_Lp_U(1-p_U)  \nonumber\\
&\quad -\dfrac{\pi_0 (1,0,1)-\pi_0 (1,0,0)}{\{1-\pi_0 (1,0,1)\}p_U+\{1-\pi_0 (1,0,0)\}(1-p_U)}\{\mu_0 (1,0,1)-\mu_0 (1,0,0)\}(1-p_L)p_U(1-p_U)\nonumber\\
&\quad +\dfrac{\pi_0 (0,1,1)-\pi_0 (0,1,0)}{\{1-\pi_0 (0,1,1)\}p_U+\{1-\pi_0 (0,1,0)\}(1-p_U)}\{\mu_0 (0,1,1)-\mu_0 (0,1,0)\}p_Lp_U(1-p_U) \nonumber\\
&\quad +\dfrac{\pi_0 (0,0,1)-\pi_0 (0,0,0)}{\{1-\pi_0 (0,0,1)\}p_U+\{1-\pi_0 (0,0,0)\}(1-p_U)}\{\mu_0 (0,0,1)-\mu_0 (0,0,0)\}(1-p_L)p_U(1-p_U)
\end{align*}

\subsection*{Derivation of \eqref{NI_sde}}

We follow a similar approach as that used in the derivation of \eqref{NI_cde}. By definition, the non-identification error is given by
\begin{align*}
    \{\mbox{SDE}^{a_D}_{obs}-\mbox{SDE}^{a_D}_0\} &=\{\tilde\psi(a_Y=1,a_D)-\tilde\psi(a_Y=0,a_D)\}  \notag\\ 
        &\quad -\{\psi_0(a_Y=1,a_D)-\psi_0(a_Y=0,a_D)\} 
\end{align*}
where 
\begin{align}
&\psi_0(a_Y=1, a_D)-\psi_0(a_Y=0, a_D)  \nonumber\\
&=\mu_0(1,1,1)\{1-\pi_0(a_D,1,1)\}p_Lp_U  \notag \\
  &\quad + \mu_0(1,1,0)\{1-\pi_0(a_D,1,0)\}p_L(1-p_U) \notag \\
  &\quad + \mu_0(1,0,1)\{1-\pi_0(a_D,0,1)\}(1-p_L)p_U  \notag \\
  &\quad + \mu_0(1,0,0)\{1-\pi_0(a_D,0,0)\}(1-p_L)(1-p_U) \notag \\ 
  &\quad -\mu_0(0,1,1)\{1-\pi_0(a_D,1,1)\}p_Lp_U  \notag \\
  &\quad - \mu_0(0,1,0)\{1-\pi_0(a_D,1,0)\}p_L(1-p_U) \notag \\
  &\quad - \mu_0(0,0,1)\{1-\pi_0(a_D,0,1)\}(1-p_L)p_U  \notag \\
  &\quad - \mu_0(0,0,0)\{1-\pi_0(a_D,0,0)\}(1-p_L)(1-p_U)   \label{app_sde_correct}
\end{align}

We next derive a suitable expression for $\tilde{\psi}(a_Y=1,a_D)-\tilde{\psi}(a_Y=0,a_D)$, where recall that 
\begin{equation*}
\tilde\psi(a_Y,a_D) = \sum_{l}\Pr(Y=1|A=a_Y, D=0, L=l) 
    \Pr(D=0|A=a_D, L=l) \Pr(L=l) 
\end{equation*}
To do so, it will be convenient to express $\pi(a,l):=\Pr(D=1|A=a, L=l)$ in terms of $\pi_0(a,l,u)$. By standard probability laws and the independence assumptions encoded in the DAG,
\begin{align}
\pi(a,l) &=\sum_{u}\Pr(D=1|A=a, L=l, U=u) \Pr(U=u|A=a, L=l) \nonumber\\
 &=\sum_{u}\Pr(D=1|A=a, L=l, U=u) \dfrac{\Pr(L=l,U=u|A=a)}{\sum_{u}\Pr(L=l,U=u|A=a)} \nonumber\\
 &=\dfrac{\sum_{u} \pi_0 (a,l,u)\Pr(L=l,U=u)}{\sum_{u}\Pr(L=l,U=u)} \nonumber \\
  &= \sum_{u} \pi_0 (a,l,u)\Pr(U=u) \label{app_pi_withoutu}
\end{align}

Plugging \eqref{app_mu_withoutu} and \eqref{app_pi_withoutu} and into the expression for $\tilde\psi(a_Y,a_D)$, 
\begin{align}
&\tilde\psi(a_Y=1, a_D) - \tilde\psi(a_Y=0, a_D) \notag\\
 &=\dfrac{\mu_0 (1,1,1)\{1-\pi_0 (1,1,1)\}p_U + \mu_0 (1,1,0)\{1-\pi_0 (1,1,0)\}(1-p_U)}{\{1-\pi_0 (1,1,1)\}p_U + \{1-\pi_0 (1,1,0)\}(1-p_U)}  \nonumber\\
 &\quad\quad \times \left[\{1-\pi_0 (a_D,1,1)\}p_U + \{1-\pi_0 (a_D,1,0)\}(1-p_U) \right]p_L \nonumber\\
 &+\dfrac{\mu_0 (1,0,1)\{1-\pi_0 (1,0,1)\}p_U + \mu_0 (1,0,0)\{1-\pi_0 (1,0,0)\}(1-p_U)}{\{1-\pi_0 (1,0,1)\}p_U + \{1-\pi_0 (1,0,0)\}(1-p_U)}  \nonumber\\
 &\quad\quad \times\left[\{1-\pi_0 (a_D,0,1)\}p_U + \{1-\pi_0 (a_D,0,0)\}(1-p_U)\right] (1-p_L) \nonumber\\  
 &-\dfrac{\mu_0 (0,1,1)\{1-\pi_0 (0,1,1)\}p_U + \mu_0 (0,1,0)\{1-\pi_0 (0,1,0)\}(1-p_U)}{\{1-\pi_0 (0,1,1)\}p_U + \{1-\pi_0 (0,1,0)\}(1-p_U)}  \nonumber\\
 &\quad\quad \times\left[\{1-\pi_0 (a_D,1,1)\}p_U + \{1-\pi_0 (a_D,1,0)\}(1-p_U) \right]  p_L \nonumber\\
 &-\dfrac{\mu_0 (0,0,1)\{1-\pi_0 (0,0,1)\}p_U + \mu_0 (0,0,0)\{1-\pi_0 (0,0,0)\}(1-p_U)}{\{1-\pi_0 (0,0,1)\}p_U + \{1-\pi_0 (0,0,0)\}(1-p_U)}  \nonumber\\
 &\quad\quad \times\left[\{1-\pi_0 (a_D,0,1)\}p_U + \{1-\pi_0 (a_D,0,0)\}(1-p_U) \right] (1-p_L) \nonumber  
\end{align}

which can be re-expressed as
\begin{align}
&\tilde\psi(a_Y=1, a_D) - \tilde\psi(a_Y=0, a_D) \notag\\
 &=\dfrac{\{1-\pi_0 (1,1,1)\}[\{1-\pi_0 (a_D,1,1)\}p_U + \{1-\pi_0 (a_D,1,0)\}(1-p_U)]}{\{1-\pi_0 (1,1,1)\}p_U + \{1-\pi_0 (1,1,0)\}(1-p_U)}\mu_0 (1,1,1)p_Lp_U \nonumber\\
  &\quad+\dfrac{\{1-\pi_0 (1,1,0)\}[\{1-\pi_0 (a_D,1,1)\}p_U + \{1-\pi_0 (a_D,1,0)\}(1-p_U)]}{\{1-\pi_0 (1,1,1)\}p_U + \{1-\pi_0 (1,1,0)\}(1-p_U)}\mu_0 (1,1,0)p_L(1-p_U)  \nonumber\\
  &\quad+\dfrac{\{1-\pi_0 (1,0,1)\}[\{1-\pi_0 (a_D,0,1)\}p_U + \{1-\pi_0 (a_D,0,0)\}(1-p_U)]}{\{1-\pi_0 (1,0,1)\}p_U + \{1-\pi_0 (1,0,0)\}(1-p_U)}\mu_0 (1,0,1)(1-p_L)p_U \nonumber\\
  &\quad+\dfrac{\{1-\pi_0 (1,0,0)\}[\{1-\pi_0 (a_D,0,1)\}p_U + \{1-\pi_0 (a_D,0,0)\}(1-p_U)]}{\{1-\pi_0 (1,0,1)\}p_U + \{1-\pi_0 (1,0,0)\}(1-p_U)}\mu_0 (1,0,0)(1-p_L)(1-p_U)  \nonumber\\ 
  &\quad-\dfrac{\{1-\pi_0 (0,1,1)\}[\{1-\pi_0 (a_D,1,1)\}p_U + \{1-\pi_0 (a_D,1,0)\}(1-p_U)]}{\{1-\pi_0 (0,1,1)\}p_U + \{1-\pi_0 (0,1,0)\}(1-p_U)}\mu_0 (0,1,1)p_Lp_U \nonumber\\
  &\quad-\dfrac{\{1-\pi_0 (0,1,0)\}[\{1-\pi_0 (a_D,1,1)\}p_U + \{1-\pi_0 (a_D,1,0)\}(1-p_U)]}{\{1-\pi_0 (0,1,1)\}p_U + \{1-\pi_0 (0,1,0)\}(1-p_U)}\mu_0 (0,1,0)p_L(1-p_U)  \nonumber\\
  &\quad-\dfrac{\{1-\pi_0 (0,0,1)\}[\{1-\pi_0 (a_D,0,1)\}p_U + \{1-\pi_0 (a_D,0,0)\}(1-p_U)]}{\{1-\pi_0 (0,0,1)\}p_U + \{1-\pi_0 (0,0,0)\}(1-p_U)}\mu_0 (0,0,1)(1-p_L)p_U \nonumber\\
  &\quad-\dfrac{\{1-\pi_0 (0,0,0)\}[\{1-\pi_0 (a_D,0,1)\}p_U + \{1-\pi_0 (a_D,0,0)\}(1-p_U)]}{\{1-\pi_0 (0,0,1)\}p_U + \{1-\pi_0 (0,0,0)\}(1-p_U)}\mu_0 (0,0,0)(1-p_L)(1-p_U)   \label{app_sde_incorrect}
\end{align}

Plugging  \eqref{app_sde_correct} and \eqref{app_sde_incorrect} into the expression for $\{\mbox{SDE}^{a_D}_{obs}-\mbox{SDE}^{a_D}_0\}$, 
\begin{align*}
&\{\mbox{SDE}^{a_D}_{obs}-\mbox{SDE}^{a_D}_0\} \notag\\
&=
\left[\dfrac{\{1-\pi_0 (1,1,1)\}[\{1-\pi_0 (a_D,1,1)\}p_U + \{1-\pi_0 (a_D,1,0)\}(1-p_U)]}{\{1-\pi_0 (1,1,1)\}p_U + \{1-\pi_0 (1,1,0)\}(1-p_U)}-\{1-\pi_0(a_D,1,1)\}\right]\\
    &\quad\quad\times\mu_0 (1,1,1)p_Lp_U \nonumber\\
&\quad+\left[\dfrac{\{1-\pi_0 (1,1,0)\}[\{1-\pi_0 (a_D,1,1)\}p_U + \{1-\pi_0 (a_D,1,0)\}(1-p_U)]}{\{1-\pi_0 (1,1,1)\}p_U + \{1-\pi_0 (1,1,0)\}(1-p_U)}-\{1-\pi_0(a_D,1,0)\}\right]\\
    &\quad\quad\times\mu_0 (1,1,0)p_L(1-p_U)  \nonumber\\
&\quad+\left[\dfrac{\{1-\pi_0 (1,0,1)\}[\{1-\pi_0 (a_D,0,1)\}p_U + \{1-\pi_0 (a_D,0,0)\}(1-p_U)]}{\{1-\pi_0 (1,0,1)\}p_U + \{1-\pi_0 (1,0,0)\}(1-p_U)}-\{1-\pi_0(a_D,0,1)\}\right]\\
    &\quad\quad\times\mu_0 (1,0,1)(1-p_L)p_U \nonumber\\
&\quad+\left[\dfrac{\{1-\pi_0 (1,0,0)\}[\{1-\pi_0 (a_D,0,1)\}p_U + \{1-\pi_0 (a_D,0,0)\}(1-p_U)]}{\{1-\pi_0 (1,0,1)\}p_U + \{1-\pi_0 (1,0,0)\}(1-p_U)}-\{1-\pi_0(a_D,0,0)\}\right]\\
    &\quad\quad\times\mu_0 (1,0,0)(1-p_L)(1-p_U)  \\
&\quad-\left[\dfrac{\{1-\pi_0 (0,1,1)\}[\{1-\pi_0 (a_D,1,1)\}p_U + \{1-\pi_0 (a_D,1,0)\}(1-p_U)]}{\{1-\pi_0 (0,1,1)\}p_U + \{1-\pi_0 (0,1,0)\}(1-p_U)}-\{1-\pi_0(a_D,1,1)\}\right]\\
    &\quad\quad\times\mu_0 (0,1,1)p_Lp_U \nonumber\\
&\quad-\left[\dfrac{\{1-\pi_0 (0,1,0)\}[\{1-\pi_0 (a_D,1,1)\}p_U + \{1-\pi_0 (a_D,1,0)\}(1-p_U)]}{\{1-\pi_0 (0,1,1)\}p_U + \{1-\pi_0 (0,1,0)\}(1-p_U)}-\{1-\pi_0(a_D,1,0)\}\right]\\
    &\quad\quad\times\mu_0 (0,1,0)p_L(1-p_U)  \nonumber\\
&\quad-\left[\dfrac{\{1-\pi_0 (0,0,1)\}[\{1-\pi_0 (a_D,0,1)\}p_U + \{1-\pi_0 (a_D,0,0)\}(1-p_U)]}{\{1-\pi_0 (0,0,1)\}p_U + \{1-\pi_0 (0,0,0)\}(1-p_U)}-\{1-\pi_0(a_D,0,1)\}\right]\\
    &\quad\quad\times\mu_0 (0,0,1)(1-p_L)p_U \nonumber\\
&\quad-\left[\dfrac{\{1-\pi_0 (0,0,0)\}[\{1-\pi_0 (a_D,0,1)\}p_U + \{1-\pi_0 (a_D,0,0)\}(1-p_U)]}{\{1-\pi_0 (0,0,1)\}p_U + \{1-\pi_0 (0,0,0)\}(1-p_U)}-\{1-\pi_0(a_D,0,0)\}\right]\\
    &\quad\quad\times\mu_0 (0,0,0)(1-p_L)(1-p_U)  
\end{align*}

Upon algebraic simplification,
\begin{align*}
&\{\mbox{SDE}^{a_D}_{obs}-\mbox{SDE}^{a_D}_0\} \notag\\
&=\dfrac{\{1-\pi_0 (1,1,1)\}\{1-\pi_0 (a_D,1,0)\}-\{1-\pi_0 (a_D,1,1)\}\{1-\pi_0 (1,1,0)\}}{\{1-\pi_0 (1,1,1)\}p_U + \{1-\pi_0 (1,1,0)\}(1-p_U)}\nonumber\\
    &\quad\quad\quad \times\left\{\mu_0 (1,1,1) - \mu_0 (1,1,0)\right\}p_Lp_U(1-p_U) \nonumber\\
&\quad+\dfrac{\{1-\pi_0 (1,0,1)\}\{1-\pi_0 (a_D,0,0)\}-\{1-\pi_0 (a_D,0,1)\}\{1-\pi_0 (1,0,0)\}}{\{1-\pi_0 (1,0,1)\}p_U + \{1-\pi_0 (1,0,0)\}(1-p_U)}\nonumber\\
    &\quad\quad\quad \times\left\{\mu_0 (1,0,1) - \mu_0 (1,0,0)\right\}(1-p_L)p_U(1-p_U) \nonumber\\
&\quad-\dfrac{\{1-\pi_0 (0,1,1)\}\{1-\pi_0 (a_D,1,0)\}-\{1-\pi_0 (a_D,1,1)\}\{1-\pi_0 (0,1,0)\}}{\{1-\pi_0 (0,1,1)\}p_U + \{1-\pi_0 (0,1,0)\}(1-p_U)}\nonumber\\
    &\quad\quad\quad \times\left\{\mu_0 (0,1,1) - \mu_0 (0,1,0)\right\}p_Lp_U(1-p_U) \nonumber\\
&\quad-\dfrac{\{1-\pi_0 (0,0,1)\}\{1-\pi_0 (a_D,0,0)\}-\{1-\pi_0 (a_D,0,1)\}\{1-\pi_0 (0,0,0)\}}{\{1-\pi_0 (0,0,1)\}p_U + \{1-\pi_0 (0,0,0)\}(1-p_U)}\nonumber\\
    &\quad\quad\quad \times\left\{\mu_0 (0,0,1) - \mu_0 (0,0,0)\right\}(1-p_L)p_U(1-p_U) \nonumber\\
\end{align*}

\section*{Web Appendix D. Near violations of positivity condition}

We discussed that even in settings where the positivity condition for competing events (6) holds, near violations of positivity may occur, which in turn inflate the variance in (14) compared to that in (17). To see this more explicitly, in our simple setting, both the estimators (12) and (15) can be written as
\begin{align}
\hat E\left[Y\times W(A=1)|A=1\right]-\hat E\left[Y\times W(A=0)|A=0\right]. 
\end{align}
differing only by the form of the weight $W(A)$.  For (12), $W(A)=\frac{I(D=0)}{\{1- \tilde{\pi}(A, L;\hat{\tilde{\beta}})\}}$  while, for (15), $W(A)=\frac{\{1- \tilde{\pi}(a_D, L;\hat{\tilde{\beta}})\}}{\{1- \tilde{\pi}(A, L;\hat{\tilde{\beta}})\}}$.  In the case of a near positivity violation for a particular joint stratum of $A,L$, an individual in this stratum who does \textsl{not} experience the competing event (an unusual individual by the premise of a near positivity violation in this stratum), will necessarily have an extremely large weight value $W(A)$ for the IPCW estimator (12) (the denominator will be close to zero while the numerator is, by definition, one).  By  contrast, even for such an ``extreme'' individual, the form of $W(A)$ for the estimator (15) is inherently ``stabilized'' (the denominator will be close to zero but the numerator is something between 0 and 1 and may also be very close to zero).  In turn, in this setting, we would expect the variance (14) to be larger than (17).

\section*{Web Appendix E. Detailed description of the dataset structure \label{time-varying}}

Let $Y_k$ and $D_k$ denote indicators of death due to prostate cancer and a competing event (death due to cardiovascular failure) by follow up month $k=0,\ldots,K+1=50$, with $Y_0=D_0=0$, indicating that all individuals are alive at baseline. To represent the history of a random variable, we use overbars, such as $\bar{Y}_k=(Y_1, Y_2, ..., Y_k)$. In this study no individual is lost to follow-up prior to 50 months such that, given no measurement error, we fully observe, ($L,A,\overline{Y}_{K+1},\overline{D}_{K+1}$) where $A$ is treatment as defined in the main text and $L$ are measured baseline covariates (see Section 5 of the main text for details).   

As \citet{young2020causal} discussed, the counterfactual risk difference at $k$ under a hypothetical intervention that eliminates competing events,
\begin{align}
    \mbox{CDE}_{k,0} \equiv \psi_{k,0}(a=1,\bar d=0)-\psi_{k,0}(a=0,\bar d=0), \label{cdecontrast_survival}
\end{align}
where 
\begin{align}
    \psi_{k,0}(a,\bar d=0) \equiv \Pr(Y_k^{a,\bar d=0}), \label{cde_survival}
\end{align}
is a controlled direct effect \citep{robins1992identifiability}. \citet{young2020causal} showed that under certain conditions, the IPCW estimator for \eqref{cde_survival} is unbiasedly estimated by
\begin{align}
    \sum_{j=0}^{k-1} \frac{\sum_{i=1}^n Y_{j+1, i}(1-Y_{j,i})W_{j,i}(\hat\eta)I(A_i=a)}{\sum_{i=1}^n (1-Y_{j,i})W_{j,i}(\hat\eta)I(A_i=a)}\prod_{s=0}^{j-1}\left[1-\frac{\sum_{i=1}^n Y_{s+1, i}(1-Y_{s,i})W_{s,i}(\hat\eta)I(A_i=a)}{\sum_{i=1}^n (1-Y_{s,i})W_{s,i}(\hat\eta)I(A_i=a)}\right],\label{ipw_cde}
\end{align}
where 
\begin{align}
    W_{k,i}(\hat\eta)=\frac{I(D_{k+1, i}=0)}{\prod_{j=0}^k [1-\Pr(D_{k+1}=1|\bar D_k=\bar Y_k=0, L_i, A=a; \hat\eta)]}, \label{weight_cde}
\end{align}
with $\Pr(D_{k+1}=0|\bar D_k=\bar Y_k=0, L_i, A=a; \eta)$ a parametric model for the competing event hazard indexed by parameter vector $\eta$, with $\hat\eta$ the MLE of $\eta$. The weight \eqref{weight_cde} is an extension of the weight used in (12), generalized by taking products over time.

\citet{stensrud2022marginal} defined the separable direct effect at $k$ as
\begin{align}
    \mbox{SDE}_{k,0} \equiv \psi_{k,0}(a_Y=1, a_D)-\psi_{k,0}(a_Y=0,a_D), \label{sdecontrast_survival}
\end{align}
where 
\begin{align}
    \psi_{k,0}(a_Y,a_D) \equiv \Pr(Y_k^{a_Y,a_D}), \label{sde_survival}
\end{align}
denoting the counterfactual risk difference at $k$ under hypothetical interventions  $a_Y=1$ versus $a_Y=0$, while holding the $a_D$ component fixed. \citet{stensrud2022marginal} showed that, under certain conditions, the weighted estimator for \eqref{sde_survival} is unbiasedly estimated by
\begin{align}
    \sum_{j=0}^{k-1} \frac{\sum_{i=1}^n Y_{j+1, i}(1-Y_{j,i})(1-D_{j+1,i})W^{a_D}_{j,i}(\hat\eta)I(A_i=a_Y)}{\sum_{i=1}^n I(A_i=a_Y)}\label{ipw_sde}
\end{align}
where 
\begin{align}
    W_{k,i}^{a_D}(\hat\eta)=\frac{\prod_{j=0}^k [1-\Pr(D_{k+1}=1|\bar D_k=\bar Y_k=0, L_i, A=a_D; \hat\eta]}{\prod_{j=0}^k [1-\Pr(D_{k+1}=1|\bar D_k=\bar Y_k=0, L_i, A=a_Y; \hat\eta]} \label{weight_sde}
\end{align}
Again, the weight is an extension of the weight used in (15), generalized by taking products over time.

The model for the competing event hazard, $\Pr(D_{k+1}=1|\bar D_k=\bar Y_k=0, L=l, A=a; \eta]$, is a nuisance function shared by both weighted estimators above. We applied a pooled logistic model to estimate this nuisance function. Estimated coefficients and standard errors are shown in Web Table 4.  Details of this model are provided in the main text, Section 5.

\section*{Web Appendix F. Approximations under rare-disease assumption \label{rare disease assumption}}
Assume that $Y$ and $D$ are rare. Then, the logistic models (18) and (19) can be approximated as
\begin{align*}
&\mu_0(a,l,u; \theta) \approx \exp\{\theta_0 + \theta_1 a + \theta_2 l + \theta_3 al + \theta_4 u + \theta_5 au + \theta_6 lu \} \\
&\pi_0(a,l,u; \beta) \approx \exp\{\beta_0 + \beta_1 a + \beta_2 l + \beta_3 a l + \beta_4 u + \beta_5 a u + \beta_6 lu\}
\end{align*}

By plugging these models into the formulas derived in Web Appendix B, we can get approximated versions or errors under the parametric assumption and rare-disease assumption. For example, the estimand error \eqref{EstimandError_cde0sde0} can be approximated as,
\begin{align*}
    &\{\exp(\theta_0 + \theta_1 + \theta_2 + \theta_3 + \theta_4 + \theta_5 + \theta_6) - \exp(\theta_0 + \theta_2 + \theta_4 + \theta_6)\} \times \\
    &\quad\quad \exp(\beta_0 + \beta_1 a_D + \beta_2 + \beta_3 a_D + \beta_4 + \beta_5 a_D + \beta_6)p_Lp_U \\
    &\quad +\{\exp(\theta_0 + \theta_1 + \theta_2 + \theta_3 ) - \exp(\theta_0 + \theta_2)\}\exp(\beta_0 + \beta_1 a_D + \beta_2 + \beta_3 a_D)p_L(1-p_U) \\
    &\quad +\{\exp(\theta_0 + \theta_1 + \theta_4 + \theta_5) - \exp(\theta_0 + \theta_4)\} \exp(\beta_0 + \beta_1 a_D + \beta_4 + \beta_5 a_D)(1-p_L)p_U \\
    &\quad +\{\exp(\theta_0 + \theta_1) - \exp(\theta_0)\}\exp(\beta_0 + \beta_1 a_D)(1-p_L)(1-p_U) \\
    &=\{\exp(\theta_1 + \theta_3 + \theta_5 ) - 1\}\exp(\theta_0 + \theta_2 + \theta_4 + \theta_6 + \beta_0 + \beta_1 a_D + \beta_2 + \beta_3 a_D + \beta_4 + \beta_5 a_D + \beta_6)p_Lp_U \\
    &\quad + \{\exp(\theta_1 + \theta_3 ) - 1\}\exp(\theta_0 + \theta_2 + \beta_0 + \beta_1 a_D + \beta_2 + \beta_3 a_D)p_L(1-p_U) \\
    &\quad + \{\exp(\theta_1 + \theta_5 ) - 1\}\exp(\theta_0 + \theta_4 + \beta_0 + \beta_1 a_D + \beta_4 + \beta_5 a_D )(1-p_L)p_U \\
    &\quad + \{\exp(\theta_1 ) - 1\}\exp(\theta_0 +\beta_0 + \beta_1 a_D )(1-p_L)(1-p_U)
\end{align*}

Consider the ratio of \eqref{app_sde_correct} to \eqref{app_cde_correct}, rather than the difference \eqref{app_cde_correct} $-$ \eqref{app_sde_correct} which we examined in \eqref{cde-estimand-error}. Assume that $p_U=0$, implying that $\{\mbox{CDE}_{obs}-\mbox{CDE}_0\}=\{\mbox{SDE}^{a_D}_{obs}-\mbox{SDE}^{a_D}_0\}=0$. If \eqref{app_cde_correct} is nonzero, under the same parametric assumption and rare-disease assumption, the ratio of \eqref{app_sde_correct} to \eqref{app_cde_correct} is approximated as,
\begin{align*}
1-\exp(\beta_0+\beta_1 a_D)\frac{\exp(\theta_2 + \beta_2 + \beta_3 a_D)\{\exp(\theta_1 +\theta_3)-1\}p_L + \{\exp(\theta_1)-1\}(1-p_L)}{\exp(\theta_2)\{\exp(\theta_1 +\theta_3)-1\}p_L + \{\exp(\theta_1)-1\}(1-p_L)}    
\end{align*}
This function can be negative. That is,  \eqref{app_cde_correct} and \eqref{app_sde_correct}, and in turn, $\mbox{CDE}_0$ and $\mbox{SDE}^{a_D}_0$ can have different signs under some sets of coefficients. 

\newpage

\begin{table}
\caption{Ranges of parameter values for logistic models (18) and (19) in data generating scenarios for Figures 3 and 4. The intercept $\theta_0$ for the model (18) was fixed at -1. For the model (19), the intercept $\beta_0$ was set to -1 for all panels in Figure 3. $\beta_0$ was set to -9 (left panels) and -6 (right panels) in Figure \ref{fig:EstimandError_rare}. In Figure 4, $\beta_0$ was set to $-1$.
}
\begin{center}
\begin{tabular}{ l l l  }
 \hline 
 Parameter & Description & Coefficients \\
 \hline 
 $\theta_1$ & Dependence of $Y$ on $A$ & -1, -0.5, 0.5, 1 \\ 
 $\theta_2$ & Dependence of $Y$ on $L$ & -1, -0.5, 0.5, 1  \\ 
 $\theta_3$ & Dependence of $Y$ on the interaction of $(A, L)$ & -1, -0.5, 0.5, 1  \\ 
 $\theta_4$ & Dependence of $Y$ on $U$ & -1, -0.5, 0.5, 1  \\ 
 $\theta_5$ & Dependence of $Y$ on the interaction of $(A, U)$  & -1, -0.5, 0.5, 1   \\ 
 $\theta_6$ &  Dependence of $Y$ on the interaction of $(L, U)$ & -1, -0.5, 0.5, 1   \\ 
 $\beta_1$ & Dependence of $D$ on $A$ & -1, -0.5, 0.5, 1  \\ 
 $\beta_2$ & Dependence of $D$ on $L$ & -1, -0.5, 0.5, 1  \\ 
 $\beta_3$ & Dependence of $D$ on the interaction of $(A, L)$ & -1, -0.5, 0.5, 1  \\ 
 $\beta_4$ & Dependence of $D$ on $U$ & -1, -0.5, 0.5, 1  \\ 
 $\beta_5$ & Dependence of $D$ on the interaction of $(A, U)$ & -1, -0.5, 0.5, 1   \\ 
 $\beta_6$ &  Dependence of $D$ on the interaction of $(L, U)$ & -1, -0.5, 0.5, 1  \\
 \hline
\end{tabular}
\label{table:parameters}
\end{center}
\end{table}

\begin{table}[ht]
\caption{Specifications of the logistic model coefficients used in simulation study, Section 4.3. In all scenarios, $(p_L, p_U, \theta_0, \theta_1, \theta_2, \theta_3)$ were fixed at $(0.1, 0.5, -1, -2, 1, 3)$}
\label{table:simulation_coefficients}
\small
\centering
\begin{tabular}{ccccccccccccc}
  \hline
\begin{tabular}{c} Near \\ positivity \\ violation? \end{tabular}  & 
\begin{tabular}{c} Dependence of\\ $Y$ and/or $D$\\ on $U$? \end{tabular} & 
\begin{tabular}{c} Competing\\ events \\ marginally rare? \end{tabular} & 
$\theta_4$ & $\theta_5$ & $\theta_6$ &
$\beta_0$ & $\beta_1$ & $\beta_2$ & $\beta_3$ & $\beta_4$ & $\beta_5$ & $\beta_6$ \\
  \hline
No & No & Yes & 0 & 0 & 0 & -3 & 1 & 1 & -1 & 0 & 0 & 0 \\ 
  No & No & No & 0 & 0 & 0 & -1 & 1 & 1 & -1 & 0 & 0 & 0 \\ 
  No & Yes & No & 1 & -2 & 0 & -3 & 1 & 1 & -1 & 3 & 1 & 0 \\ 
  Yes & No & Yes & 0 & 0 & 0 & -10 & 1 & 16 & -1 & 0 & 0 & 0 \\ 
  Yes & No & No & 0 & 0 & 0 & -1 & 1 & 7 & -1 & 0 & 0 & 0 \\ 
  Yes & Yes & No & 1 & -2 & 0 & -3 & 1 & 7 & -1 & 3 & 1 & 0 \\ 
   \hline
\end{tabular}
\end{table}

\begin{table}[ht]
\caption{Simulation-based comparison of the variance Var($\widehat{\mbox{SDE}}^{a_D=1}_{obs}$). Expectations in the variance calculations were taken relative to the distribution over simulation runs.}
\centering
\begin{tabular}{cccrr}
  \hline
\begin{tabular}{c} Near \\ positivity \\ violation? \end{tabular}  & 
\begin{tabular}{c} Dependence of\\ $Y$ and/or $D$\\ on $U$? \end{tabular} & 
\begin{tabular}{c} Competing\\ events \\ marginally rare? \end{tabular} & 
\begin{tabular}{l} Var($\widehat{\mbox{SDE}}^{a_D=1}_{obs}$)
\end{tabular} \\
  \hline
No & No & Yes &  $5 \times10^{-6}$ \\ 
  No & No & No & $3 \times10^{-6}$  \\ 
  No & Yes & No & $3 \times10^{-6}$ \\ 
  Yes & No & Yes & $4 \times10^{-6}$ \\ 
  Yes & No & No & $2 \times10^{-6}$ \\ 
  Yes & Yes & No & $2 \times10^{-6}$ \\ 
   \hline
\end{tabular}
\label{table:simulation_results_sde1}
\end{table}

\begin{table}
\caption{Regression coefficients for other-cause death by the pooled logistic model fitted to the prostate cancer data used in Section 5.}
\centering
\begin{tabular}{lrr}
  \hline
 & Estimate & Std. Error \\ 
  \hline
Intercept & -5.12 & 0.71 \\ 
  Month & -0.02 & 0.02 \\ 
  Month$^2$ & 0.00 & 0.00 \\ 
  Activity level (normal activity) & -0.28 & 0.30 \\ 
  Age (60-75) & 0.70 & 0.60 \\ 
  Age ($\geq$ 75) & 1.09 & 0.61 \\ 
  History of cardiovascular disease & 0.33 & 0.30 \\ 
  Serum hemoglobin ($<$12 g/100ml) & 0.60 & 0.24 \\ 
  Treatment (5.0 mg estrogen) & -0.14 & 0.31 \\ 
  Treatment × History of cardiovascular disease & 0.76 & 0.41 \\ 
   \hline
\end{tabular}
\label{table:pooled}
\end{table}

\newpage

\begin{figure}
\includegraphics[width=\textwidth, keepaspectratio]{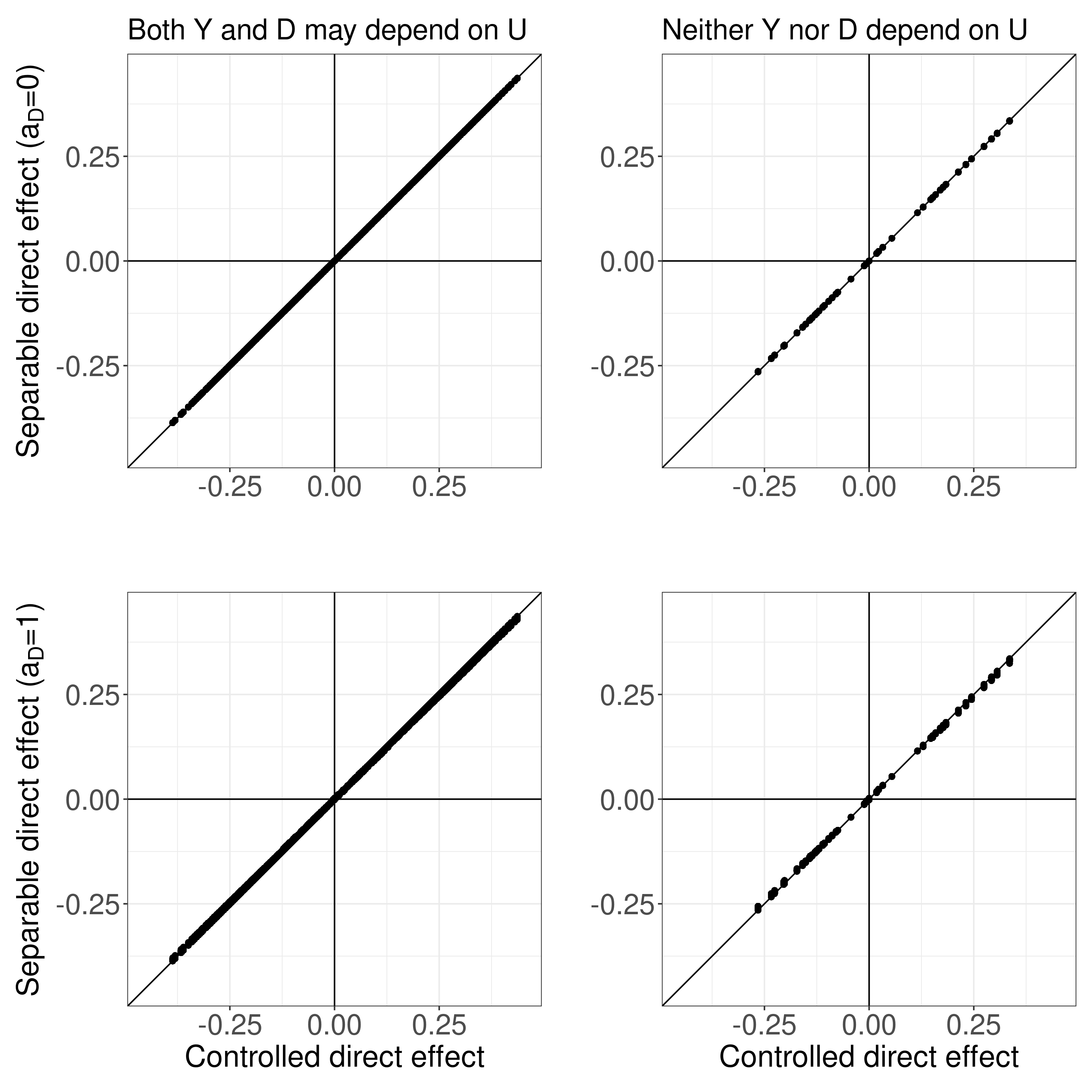}
\caption{Illustration of estimand error when the actual causal target is a separable direct effect in the scenarios where $D$ is rare (i.e., under the parameters given in Web Table \ref{table:parameters} where $\Pr(D=1|A=a,L=l,U=u) < 10\%$ for all $(a,l,u)$).}
\label{fig:EstimandError_rare}
\end{figure}
\end{document}